\documentclass[twocolumn,preprintnumbers,amsmath,amssymb]{revtex4}

\usepackage{graphicx}
\usepackage{dcolumn}
\usepackage{bm}
\usepackage{epsfig}

\begin{document}


\title{Explicit Thin-Lens Solution\\ for an Arbitrary Four by Four Uncoupled
Beam Transfer Matrix}

\author{V.Balandin}
\email{vladimir.balandin@desy.de}
\affiliation{
Deutsches Elektronen-Synchrotron DESY, Notkestrasse 85, 22607 Hamburg, Germany}
\author{S.Orlov}%
\affiliation{%
Faculty of Computational Mathematics and Cybernetics,\\
M.V.Lomonosov Moscow State University, 119991 Moscow, Russia
}%

\date{\today}

\begin{abstract}
In the design of beam transport lines, one often meets the
problem of constructing a quadrupole lens system
that will produce desired transfer matrices in both 
the horizontal and vertical planes. 
Nowadays this problem is typically approached with the help of
computer routines, but searching for the numerical solution
one has to remember that it is not proven yet
that an arbitrary four by four uncoupled beam transfer matrix
can be represented by using a finite number of drifts and quadrupoles
(representation problem) and the answer to this question is not known 
not only for more or less realistic quadrupole field models but also
for the both most commonly used approximations of quadrupole focusing, namely
thick and thin quadrupole lenses.
In this paper we make a step forward in resolving the representation problem
and, by giving an explicit solution, we prove that an arbitrary four by four 
uncoupled beam transfer matrix
actually can be obtained as a product of a finite number of thin-lenses and drifts.
\end{abstract}


\maketitle

\section{Introduction}

In the design of beam transfer lines, one often encounters 
the problem of finding a combination of quadrupole lenses 
and field free spaces (drifts) that will produce particular 
transfer matrices in both the horizontal and the vertical 
planes. Nowadays this problem is typically approached 
with the help of computer routines which minimize the 
deviations from the desired matrices as function of the 
quadrupole strengths, lengths and distances between them.
Although very sophisticated software became available for 
these purposes during the past decades, there is an important 
theoretical question which has not been answered yet and 
whose answer could affect the strategy and efficiency of 
numerical computations. Searching for a numerical solution,
one has to remember that it is not proven yet that an arbitrary 
four by four uncoupled beam transfer matrix can be represented 
by using a finite number of drifts and quadrupoles
(representation problem) and the answer to this question is 
not known not only for more or less realistic quadrupole 
field models but also for the both most commonly used 
approximations of quadrupole focusing, namely thick 
and thin quadrupole lenses.

In this paper we make a step forward in resolving the 
representation problem and prove that an arbitrary four 
by four uncoupled beam transfer matrix actually can be 
obtained as a product of a finite number of thin-lenses and 
drifts. Even though our proof uses more thin lenses than 
probably needed, we believe that the solution provided is 
not only of theoretical interest, but could also find some 
practical applications because it uses explicit analytical 
formulas connecting thin-lens parameters with the elements 
of the input beam transfer matrix. 

Though the thin-lens kick is the simplest model of the 
quadrupole focusing, its role in accelerator physics can 
hardly be overestimated. The thin-lens quadrupole approximation 
reveals the analogy between light optics and charged 
particle optics and, if one takes into account difficulties of 
analytical manipulations with the next by complexity
thick-lens quadrupole model ~\cite{Regenstreif_1, Regenstreif_2},
is an indispensable tool for 
understanding principles and limitations of the already 
available optics modules and for development of the new 
optics solutions (see, as good examples, 
papers ~\cite{BrownServranckx, MontagueRuggiero, Zotter, Napoly, dAmigoGuignard}).

The paper by itself is organized as follows. In Sec. II we 
introduce all needed notations and give the lower bound on 
the number of drifts and lenses which are required for a 
solution of the representation problem by providing an 
example of a matrix which cannot be obtained using five 
thin lenses and five independently variable drift spaces. This 
result is somewhat unexpected and up to some extent contradicts 
a rather widespread opinion that the typical problem 
can be solved by taking a number of parameters equal to the 
number of constraints available. We see that although 
the four by four uncoupled beam transfer matrix has only 
6 degrees of freedom, there are matrices which cannot be 
represented not only by three thin lenses and three drifts 
(six parameters), but also by five thin lenses and five drifts 
(ten parameters). This example, the example provided by the 
matrix (\ref{i2_gb_ex2}), other of our attempts (though omitted in this 
paper) to find thin-lens decompositions for particular beam 
transfer matrices and the properties of the explicit solution 
given below in this paper, lead us to the conjecture that in 
order to represent an arbitrary four by four uncoupled beam 
transfer matrix one needs at least six thin lenses if the 
distances between them can be varied (independently or 
not) or at least seven thin lenses if this variation is not 
allowed.

In Sec. III we prove that an arbitrary four by four 
uncoupled beam transfer matrix can be obtained as a 
product of a finite number of thin-lenses and drifts by 
giving an explicit solution of the thin-lens representation 
problem which uses equally spaced thin lenses. The core 
idea of our approach is the representation of the matrix of 
the thin-lens multiplet as a product of elementary $P$ matrices
(the definition and the properties of the matrix $P$ can be 
found in Appendix A) with subsequent reduction of the 
initial 2D problem to two independent 1D problems. We 
use in this section the equally spaced thin-lens system 
because it allows one to make such a reduction with a 
minimum of technical details. The solution obtained 
utilizes 13 lenses if the spacing between them is fixed 
beforehand and 12 lenses if this distance can be used as 
an additional parameter. Thus, it uses six more lenses than 
the minimal number stated in our conjecture, but the 
setting of these six lenses depends only on the distance 
between lenses and therefore does not depend (at least 
directly) on the particular input beam transfer matrix.

In Sec. IV we consider the case of arbitrarily spaced thin
lenses. First, we show that the solution of the representation 
problem presented in the previous section is still valid 
after some minor modifications. Next we study in greater 
detail the ways to transform the matrix of the drift-lens 
system to the product of the elementary $P$ matrices (see formulas 
(\ref{b2})-(\ref{b6_2}) and (\ref{ff_1})-(\ref{ff_6}) below).
The representation of the matrix of the thin-lens multiplet as a product 
of elementary $P$ matrices (together with the multiplication formula (\ref{e1_r3}))
is a useful new tool for the analytical study 
of the properties of thin-lens systems. It also gives some 
clarification of the question why the role of the variable 
drift spaces and the role of the variable lens strengths
are different when they are used as fitting parameters.

This paper is mostly a theoretical paper and its main 
purpose is to turn the common believe that an arbitrary four 
by four uncoupled beam transfer matrix can be obtained as 
a product of a finite number of thin lenses and drifts into 
proven scientific fact. Still, both, the developed new technique 
for the analytical study of the properties of thin-lens 
multiplets and the explicit thin-lens solution presented in 
this paper, are of independent interest. To illustrate that, in 
Appendix B we apply our $P$ matrix approach to the study 
of four-lens beam magnification telescopes and find new, 
previously unknown analytical solutions for this important 
optics module. In Appendix C, we apply the explicit 
solution developed in this paper to the design of a beam
line which allows an independent scan of horizontal and 
vertical phase advances while preserving the entrance and 
exit matching conditions for the Twiss parameters.

Besides that thin-lens blocks with decoupled transverse 
actions introduced in this paper are another point of general 
interest. Although the idea of decoupled tuning knobs by 
itself is not new in the field of accelerator physics (see, for 
example, ~\cite{Roser, WalkerIrwinWoodley}), 
our approach is new and is not based on an iterative usage of small
steps in the lens strengths obtained at each iteration by linearization.

\section{Statement of the problem and preliminary considerations}

Let $M$ be an arbitrary four by four uncoupled beam 
transfer matrix and let the two by two symplectic matrices 
$M_x$ and $M_y$ be its horizontal and vertical focusing blocks,
respectively. Let us denote by $Q(g)$ the transfer matrix of 
the one-dimensional thin lens of strength $g$ and by $D(l)$ 
the transfer matrix of the one-dimensional drift space of length $l$: 

\noindent
\begin{eqnarray}
Q(g) \,=\, 
\left(
\begin{array}{rr}
1 &  0 \\
g &  1      
\end{array}
\right),
\;\;\;\;\;
D(l) \,=\, 
\left(
\begin{array}{rr}
1 &  l \\
0 &  1      
\end{array}
\right).
\label{i1}
\end{eqnarray}

\noindent
The problem of representation of the matrix $M$
by a thin-lens system can then be written as

\noindent
\begin{eqnarray}
D(l_n)\, Q(\pm g_n) \cdot \ldots \cdot D(l_1)\, Q(\pm g_1) \,=\, M_{x, y},
\label{i2}
\end{eqnarray}

\noindent
where (here and later on) one has to take the upper sign in 
the combinations $\pm$ and $\mp$ together with the index $x$ and the 
lower sign together with the index $y$.

Note that the drift-lens system presented on the left-hand 
side of Eq. (\ref{i2}) consists of equal numbers of drifts and 
lenses and the first element which the beam sees during 
its passage is a thin-lens. 
Alternatively, one can consider equation

\noindent
\begin{eqnarray}
Q(\pm g_n)\,D(l_n) \cdot \ldots \cdot  Q(\pm g_1)\,D(l_1) \,=\, M_{x, y},
\label{i2_sfd}
\end{eqnarray}

\noindent
where the first element is a drift space, or one can use the 
drift-lens system with a nonequal number of drifts and 
lenses which starts and ends with a drift (or a lens), but 
for the moment this is not important.

There are many unanswered questions related to Eq. (\ref{i2}),
the most interesting for us in this paper is the following:
given a matrix $M$, does there exist a number $n$ such that 
these equations have a solution? If the answer to this 
question is positive, could the number $n$ be chosen
independently from the input matrix $M$ and, if it is also 
possible, what is the minimal $n$ required?

From a mathematical point of view, Eq. (\ref{i2}) is a system 
of eight polynomial equations in $2 n$ unknowns and for any 
polynomial system considered over an algebraically closed 
field of complex numbers there is an algorithmic way to 
answer the question if this system has infinitely many 
solutions or has a finite number of solutions, or has no 
solutions at all. This can be done by transforming the 
original system to a special form called a Gr$\ddot{\mbox{o}}$bner basis 
and, very loosely speaking, is an analogue of the Gaussian 
elimination process in linear algebra ~\cite{CoxLittleOshea}.
The Gr$\ddot{\mbox{o}}$bner basis can be computed in finitely many steps
and, moreover, nowadays its calculation can be done 
with the help of symbolic manipulation programs like
MATHEMATICA and MAPLE.

Unfortunately, we are interested in the real solutions 
of Eq. (\ref{i2}) constrained additionally by the requirements
for the drift lengths to be nonnegative
and therefore we cannot use all benefits provided by the Gr$\ddot{\mbox{o}}$bner basis theory.
Nevertheless, the use of the Gr$\ddot{\mbox{o}}$bner basis approach, 
although it did not help us to solve the problem in general, 
it was very useful in providing examples of particular matrices
which cannot be obtained using a certain number of thin lenses and drift spaces.
For example,  using the  
Gr$\ddot{\mbox{o}}$bner basis technique,
it is possible to prove that the matrix $M$ with

\noindent
\begin{eqnarray}
M_x \,=\,M_y\,=\,
\left(
\begin{array}{rr}
 1 & 0\\
-1 & 1
\end{array}
\right)
\label{i2_gb}
\end{eqnarray}

\noindent
cannot be represented by five thin lenses and five
variable drift spaces starting either from a lens 
like in Eq. (\ref{i2})
or from a drift
like in Eq. (\ref{i2_sfd}).

This example, the example provided by the matrix (\ref{i2_gb_ex2}), 
many other of our attempts to study the representation problem  
for particular beam transfer matrices,
and the properties of the explicit solution given below in this paper lead 
us to the conjecture that in order to be able to represent 
an arbitrary four by four uncoupled beam transfer matrix 
one needs at least six thin lenses if the distances between them can
be varied (independently or not) or at least seven thin lenses 
with nonzero drift spaces between them if this variation is not allowed.

To finish this section, let us note that
in the above discussions we made no use of the fact that we are interested
not in the general system of polynomial equations, but only in the polynomial
system produced by a product of matrices with simple inversion
properties:

\noindent
\begin{eqnarray}
Q^{-1}(g)\,=\,Q(-g),
\;\;\;\;\;
D^{-1}(l)\,=\,D(-l).
\label{i2_invp}
\end{eqnarray}
 
\noindent
Choosing some $k = 1, \ldots, n-1$
and using (\ref{i2_invp}),  
one can rewrite the system (\ref{i2}) in the equivalent form:

\noindent
\begin{eqnarray}
D(l_k)\, Q(\pm g_k) \cdot \ldots \cdot D(l_1)\, Q(\pm g_1) \,=
\nonumber
\end{eqnarray}
\noindent
\begin{eqnarray}
Q(\mp g_{k+1}) \,D(-l_{k+1}) \cdot \ldots \cdot  Q(\mp g_n) \,D(-l_n)\,
M_{x, y}.
\label{i2_do}
\end{eqnarray}

\noindent
This trick can be used for the elimination of a part of the unknowns
from the original system
by solving Eq. (\ref{i2_do}) with respect to the variables
$g_1, \ldots, g_k, l_1, \ldots, l_k$ 
or one may even think to construct an iterative 
solution method 
which could be considered as matrix version of the
method of successive elimination of unknowns 
~\cite{Napoly, ChaoIrwin}.
This method was developed especially to deal with 
the thin-lens multiplets and was
used in ~\cite{ChaoIrwin} in an attempt to characterize
all uncoupled beam transfer matrices which can be obtained by using three
thin lenses and three drift spaces.  
Unfortunately, however this approach did not give us any 
additional noticeable simplifications
in the solution of the general representation problem.

\section{Solution of 2D problem using equally spaced thin lenses}

In this section we will give an explicit solution of the thin-lens representation 
problem which uses equally spaced thin-lenses.
Instead of Eq. (\ref{i2}) or Eq. (\ref{i2_sfd}), we will consider the system

\noindent
\begin{eqnarray}
B(m_n,\, \pm g_n, \,p_n) \cdot \ldots \cdot B(m_1,\, \pm g_1,\, p_1) \,=\, M_{x, y},
\label{a1_0}
\end{eqnarray}

\noindent
where as an elementary building block we take a thin lens sandwiched
between two drift spaces

\noindent
\begin{eqnarray}
B(m, \,\pm g, \,p) \,=\,D(p)\,Q(\pm g)\,D(m).
\label{a1}
\end{eqnarray}

If the block length $\,l = m + p > 0$, then
one can represent the block transfer matrix in the form

\noindent
\begin{eqnarray}
B(m, \,\pm g, \, p) \,=\,S^{-1}(m, \,p)\,P(2 \pm l g)\,S(m,\, p),  
\label{a2}
\end{eqnarray}

\noindent
where

\noindent
\begin{eqnarray}
S(m,\,p) \,=\, \frac{1}{\sqrt{l}} 
\left(
\begin{array}{rr}
 1 &  m \\
-1 &  p      
\end{array}
\right)
\label{a3}
\end{eqnarray}

\noindent
and

\noindent
\begin{eqnarray}
P(a) \,=\, 
\left(
\begin{array}{rr}
 a &  1 \\
-1 &  0      
\end{array}
\right).
\label{a3_1}
\end{eqnarray}

\noindent
Note that the properties of the matrix $P$
(and other elementary matrices used in this paper)
can be found in Appendix A.

Let us assume that in the system (\ref{a1_0})
all $m_k$ and all $p_k$ are equal to each other, i.e., that

\noindent
\begin{eqnarray}
m_1 = \ldots = m_n = m, 
\;\;\;\; 
p_1 = \ldots = p_n = p,
\label{es0}
\end{eqnarray}

\noindent
and let $\,l = m + p > 0$.
The principle simplification that occurs in this case 
is that after the substitution of the representation (\ref{a2})
into Eq. (\ref{a1_0}) the matrices $S(m, p)$ and
$S^{-1}(m, p)$ cancel each other
and we obtain

\noindent
\begin{eqnarray}
P(2 \pm l g_n) \cdot \ldots \cdot P(2 \pm l \,g_1) \,=\, \hat{M}_{x, y},
\label{es1}
\end{eqnarray}

\noindent
where

\noindent
\begin{eqnarray}
\hat{M}_{x,y} \,=\, S(m, \,p) \,M_{x, y}\, S^{-1}(m,\, p).
\label{es2}
\end{eqnarray}

\noindent
Equations (\ref{es1}) give the dimensionless form of 
Eq. (\ref{a1_0}) and, additionally,
one sees that while the original system (\ref{a1_0}) is formed by 
the product of $2 n + 1$ interleaved thin-lens and drift matrices 
(with neighboring drifts lumped together), the system (\ref{es1}) includes
only $n+2$ matrices depending on unknowns 
(there are $n+2$ unknowns: $n$ lens strengths plus 
two variables characterizing the block length and the position of the lens
inside the block)
and $n$ of them are $P$ matrices.

Nevertheless, the system (\ref{es1}) is still too complicated to
find easily its solutions (or even to prove their existence) 
for an arbitrary matrix $M$ and with the number of lenses
$n$ equal to six or seven as required by our conjecture. 
Instead we will provide an explicit solution which
utilizes 13 lenses if the parameters $m$ and $p$ are fixed
and are independent from the input matrix $M$, 
and 12 lenses if $m$ and $p$ can be varied.
The main idea of our solution is the reduction of the 2D problem (\ref{es1})
to two independent or, more exactly, almost independent 1D problems
by constructing thin-lens blocks 
which can act in the horizontal and the vertical planes
similar to a single $P$ matrix, but whose actions for both planes
can be chosen independently.
At first we will consider a solution of the 1D problem in terms
of $P$ matrices. As the next step we will introduce a four-lens block with decoupled
transverse actions and then will give an explicit solution of the complete 2D problem.  
Besides that we will discuss the recipe for constructing lens blocks
with decoupled transverse actions with more than four lenses.

Before giving the technical details
let us consider one more example obtained with the help of the
Gr$\ddot{\mbox{o}}$bner basis technique.
Let us assume that $m$ and $p$ are fixed and let the matrix $M$ be such 
that the matrix $\hat{M}$ in (\ref{es1}) is equal to the symplectic unit matrix:

\noindent
\begin{eqnarray}
\hat{M}_x \,=\,\hat{M}_y\,=\,
\left(
\begin{array}{rr}
 0 & 1\\
-1 & 0
\end{array}
\right).
\label{i2_gb_ex2}
\end{eqnarray}

\noindent
Then this matrix $M$ can not be represented by less than seven thin lenses 
and with seven lenses there are many solutions which
geometrically can be viewed as six distinct parallel straight lines in the 
seven-dimensional
space of lens strengths.

\subsection{1D problem in terms of $P$ matrices}

According to our plan we will prove in this subsection that
every real symplectic $2 \times 2$ matrix $M = (m_{ij})$ can be
represented as a product of at most four $P$ matrices.
First, we will consider the  
case of three $P$ matrices and will find that three $P$ matrices 
are insufficient for the representation of an arbitrary $2 \times 2$ 
symplectic matrix.
Next we will switch to the case of four $P$ matrices and
will show that with four $P$ matrices a solution can always be found,
but it is always nonunique.

Let us start with the case of three $P$ matrices, i.e., from the
equation

\noindent
\begin{eqnarray}
P(z_3)\, P(z_2)\, P(z_1) \,=\,M.
\label{s1d1}
\end{eqnarray}

\noindent
This matrix equation is, in fact,
the system of the four equations for the four matrix elements 

\noindent
\begin{eqnarray}
\left\{
\begin{array}{l}
z_3 \cdot (z_1 \, z_2 - 1) - z_1 = m_{11}\\
z_2 = -m_{22}\\
z_2 \, z_3 - 1 = m_{12}\\
z_1 \, z_2 - 1 = -m_{21}
\end{array}
\right.
\label{s1d4_01}
\end{eqnarray}

\noindent
and, as it is well known, due to symplecticity of the matrices
on both sides of (\ref{s1d1}) these four equations should be
equivalent to some system consisting of three equations only.
In order to obtain such a system let us first substitute
$z_1 \, z_2 - 1 = -m_{21}$ into the first equation of the system 
(\ref{s1d4_01}) and then plug $z_2 = -m_{22}$
in the equations three and four. Because in the resulting system

\noindent
\begin{eqnarray}
\left\{
\begin{array}{l}
z_1 = -m_{11} - m_{21} \cdot z_3\\
z_2 = -m_{22}\\
m_{22} \cdot z_3 = - 1 - m_{12}\\
m_{22} \cdot z_1 = - 1 + m_{21}
\end{array}
\right.
\label{s1d4_02}
\end{eqnarray}

\noindent
the fourth equation is equal to the first equation
multiplied by $m_{22}$ minus the third equation
multiplied by $m_{21}$, it can be omitted. 
Thus we obtain that the system of the four third order 
polynomial equations (\ref{s1d4_01}) is equivalent to the system

\noindent
\begin{eqnarray}
\left\{
\begin{array}{l}
z_1 = -m_{11} - m_{21} \cdot z_3\\
z_2 = -m_{22}\\
m_{22} \cdot z_3 = - 1 - m_{12}
\end{array}
\right.
\label{s1d4}
\end{eqnarray}

\noindent
which is linear in the unknowns $z_1$, $z_2$, and $z_3$.
Moreover, this system already has a triangular form and
its solvability depends only on the solvability  of the
third equation with respect to the variable $z_3$.

Elementary analysis shows that there are three possibilities 
for the solutions of the system (\ref{s1d4}).
If $m_{22} \neq 0$, then there exists a unique solution

\noindent
\begin{eqnarray}
z_1 = \frac{m_{21} - 1}{m_{22}}, \;\;\;
z_2 = -m_{22}, \;\;\;
z_3 = -\frac{m_{12} + 1}{m_{22}}.
\label{s1d2}
\end{eqnarray}

\noindent
If $m_{22} = 0$  and $m_{21} = 1$ (i.e if
$M = -P(-m_{11})$), then there exists a one-parameter family of solutions:

\noindent
\begin{eqnarray}
z_1 + z_3 = -m_{11}, \;\;\;
z_2 = 0.
\label{s1d3}
\end{eqnarray}

\noindent
Finally, if $m_{22} = 0$  and $m_{21} \neq 1$, then there is no solution at all.

Very loosely speaking, the condition $m_{22} = 0$ defines the two-dimensional surface of
singularities in the three-dimensional space of $2 \times 2$ real symplectic matrices. 
This surface, in the next turn, contains the one-dimensional curve selected by the
additional relation $m_{21} = 1$. If the matrix $M$ (represented as a point in our 
three-dimensional space) lies outside of the surface of singularities, then a solution
for such a matrix exists and is unique. If the point representing the matrix $M$
belongs to the surface of singularities, then we either have many solutions 
or none depending on whether this point lies on the above defined 
one-dimensional curve or not.

Let us now turn our attention to the equation

\noindent
\begin{eqnarray}
P(z_4)\,P(z_3)\, P(z_2)\, P(z_1) \,=\,M,
\label{s1d5}
\end{eqnarray}

\noindent
which includes four $P$ matrices.
The equivalent to this equation system is given below:

\noindent
\begin{eqnarray}
\left\{
\begin{array}{l}
z_1 = m_{21} - (m_{11} + m_{21} \cdot z_4) \cdot z_3\\
z_2 = -m_{12} - m_{22} \cdot z_4\\
(m_{12} + m_{22} \cdot z_4) \cdot z_3 = m_{22} - 1
\end{array}
\right.
\label{s1d6}
\end{eqnarray}

\noindent
and the easiest way to obtain it is to substitute
into the system (\ref{s1d4}) the elements of the matrix $P^{-1}(z_4)\cdot M$
instead of the $m_{ij}$. 

The system (\ref{s1d6}) is not linear anymore, but still 
has a triangular form
and its solvability depends again only on the solvability  of the
third equation with respect to the variables $z_3$ and $z_4$.
Because the matrix $M$ is nondegenerated its elements
$m_{12}$ and $m_{22}$ cannot be equal to zero simultaneously
and therefore the expression $m_{12} + m_{22} \cdot z_4$ 
considered as a function of $z_4$
cannot be equal to zero in more than one point. It means
that the last equation in (\ref{s1d6}) always has 
solutions and a good way to understand their 
complete structure is to consider this equation
as the equation of a curve on the plane $(z_3, z_4)$.
If $m_{22} \cdot (m_{22}-1) \neq 0$ this curve is a hyperbola
with two separate branches, if $m_{22} = 1$ it is a degenerate
hyperbola consisting of two intersecting lines
$z_3 = 0$ and $z_4 = -m_{12}$, and, finally, if $m_{22} = 0$
we have a single straight line $z_3 = - m_{12}^{-1}$.
So we see that with the help of the four $P$ matrices a solution of our problem
can always be found and is always nonunique.

\subsection{Four-lens block with decoupled transverse actions}

Let us denote by $W^{x,y}$ the following combination
of four $P$ matrices:

\noindent
\begin{eqnarray}
W^{x,y}=P(2 \pm l g_4) P(2 \pm l g_3) P(2 \pm l g_2) P(2 \pm l g_1),
\label{es3}
\end{eqnarray}

\noindent
which in the original variables (\ref{a1_0}) includes
four thin-lenses (four-lens block).

If one chooses $\delta = \pm 1$ and if one takes

\noindent
\begin{eqnarray}
g_2 = \frac{\delta \sqrt{3}}{l},\;\;\;\;\;
g_3 = -\frac{\delta \sqrt{3}}{l},
\label{es4}
\end{eqnarray}

\noindent
then the block matrix can be written as

\noindent
\begin{eqnarray}
W^{x, y} \,=\, -\Lambda^{-1} \left(\sqrt{u^{x,y}}\right)\, 
P (w^{x,y})\,\Lambda \left(\sqrt{u^{x,y}}\right),
\label{es5}
\end{eqnarray}

\noindent
where $\Lambda(a) = \mbox{diag}(a, 1 / a)$ is a diagonal scaling matrix,

\noindent
\begin{eqnarray} 
u^{x,y} \,=\, 2 \,\mp\, \delta \sqrt{3},\;\;\;\;\;
u^x \cdot u^y \,=\,1
\label{es6_3}
\end{eqnarray}

\noindent
and

\noindent
\begin{eqnarray} 
w^x \,=\, 7 \,+ \,u^y \cdot l g_1 \,+\, 
u^x \cdot l g_4, 
\label{es6_1}
\end{eqnarray}

\noindent
\begin{eqnarray} 
w^y \,=\, 7 \,- \,u^x \cdot l g_1 \,-\, 
u^y \cdot l g_4.
\label{es6_2}
\end{eqnarray}

\noindent
Since for any given value of $w^x$ and $w^y$ Eqs.
(\ref{es6_1}) and (\ref{es6_2}) can be solved with respect to
the variables $g_1$ and $g_4$,

\noindent
\begin{eqnarray} 
g_1 = -\frac{\delta \sqrt{3}}{l}\cdot
\frac{28 \,-\, u^y \cdot w^x \,-\, u^x \cdot w^y}{24},
\label{es7}
\end{eqnarray}

\noindent
\begin{eqnarray} 
g_4 = \;\,\frac{\delta \sqrt{3}}{l}\cdot
\frac{28\,-\,u^x \cdot w^x \,-\, u^y \cdot w^y}{24},
\label{es8}
\end{eqnarray}

\noindent
the formula (\ref{es5}) gives the result which we were looking for.
Both matrices $W^x$ and $W^y$ are similar to a single $P$ matrix
(with an inessential minus sign) and both parameters $w^x$ and $w^y$
can be chosen independently,
and then the setting of the first and the last
lenses in the block is determined according to the
formulas (\ref{es7}) and (\ref{es8}).

\subsection{Reduction of 2D problem to two independent or almost independent 1D problems}

Since with four $P$ matrices we always can solve the 1D problem, let us first consider a 
combination of four blocks of the type (\ref{es5}). Using (\ref{e9}), one can show that
the total matrix of this 16 lens system can be written as follows:

\noindent
\begin{eqnarray}
W^{x, y}_4 \, W^{x, y}_3 \, W^{x, y}_2 \, W^{x, y}_1 \,=\,
\Lambda \left(a^{x,y}\right) \cdot
\nonumber
\end{eqnarray}

\noindent
\begin{eqnarray}
P \left(\hat{w}^{x,y}_4\right)
P \left(\hat{w}^{x,y}_3\right)
P \left(\hat{w}^{x,y}_2\right)
P \left(\hat{w}^{x,y}_1\right)
\Lambda \left(a^{x,y}\right),
\label{am1_0}
\end{eqnarray}

\noindent
where

\noindent
\begin{eqnarray}
a^{x,y} \,=\, \sqrt{\frac{u^{x,y}_1 u^{x,y}_3}{u^{x,y}_2 u^{x,y}_4}}
\label{am2_0}
\end{eqnarray}

\noindent
and

\noindent
\begin{eqnarray}
\hat{w}^{x,y}_1 = \frac{u^{x,y}_2 u^{x,y}_4}{u^{x,y}_3}\cdot w^{x,y}_1,
\;\;\;
\hat{w}^{x,y}_2 = \frac{u^{x,y}_3}{u^{x,y}_1 u^{x,y}_4}\cdot w^{x,y}_2,
\label{am3_0}
\end{eqnarray}

\noindent
\begin{eqnarray}
\hat{w}^{x,y}_3 = \frac{u^{x,y}_1 u^{x,y}_4}{u^{x,y}_2}\cdot w^{x,y}_3,
\;\;\;
\hat{w}^{x,y}_4 = \frac{u^{x,y}_2}{u^{x,y}_1 u^{x,y}_3}\cdot w^{x,y}_4.
\label{am4_0}
\end{eqnarray}

\noindent
Plugging this representation into Eq. (\ref{es1}) we obtain

\noindent
\begin{eqnarray}
P \left(\hat{w}^{x,y}_4\right)
P \left(\hat{w}^{x,y}_3\right)
P \left(\hat{w}^{x,y}_2\right)
P \left(\hat{w}^{x,y}_1\right) =
\nonumber
\end{eqnarray}

\noindent
\begin{eqnarray}
\Lambda^{-1} \left(a^{x,y}\right)
\hat{M}_{x,y}
\Lambda^{-1} \left(a^{x,y}\right).
\label{am1_001}
\end{eqnarray}

\noindent
Let us choose arbitrary nonnegative $m$ and $p$ with $l = m + p > 0$
and select for each four-lens block its own $\delta = \pm 1$. 
This, in accordance with formula (\ref{es4}),
gives us the setting of the eight lenses in our system 
and this completely determines the matrix on the right-hand side of
Eq. (\ref{am1_001}). As the last step we take
$\hat{w}^{x}_k$ and $\hat{w}^{y}_k$ as some solutions of two
independent 1D problems of the type (\ref{s1d5}) and define
the strengths of the remaining eight lenses 
using the formulas (\ref{am3_0}), (\ref{am4_0}), (\ref{es7}), and (\ref{es8}). 

One sees that using four blocks with decoupled transverse actions
the complete 2D problem can always be reduced to two easily solvable
independent 1D problems. But do we really need four blocks for making such a reduction?
The answer is no and the reason for this is as follows. We know that for most of the
$2 \times 2$ symplectic matrices the 1D problem can be solved with three $P$ matrices,
which means that for most of the $4 \times 4$ uncoupled beam transfer matrices 
the 2D problem can also be solved with three blocks. The problem is what to do
with the rest? Happily it turns out that by appropriate choice of the parameters
$m$ and $p$ one can always move the input matrix $M$ away from the region of
unsolvability and, if the variation of $m$ and $p$ is not allowed, 
this can be done by using only one additional thin lens.
Thus, we arrive at the solution  announced in the Introduction,
namely 13 lenses if the spacing between them is fixed 
and 12 lenses if this distance can be used as an additional parameter.
Below we will consider in detail the case of 12 lenses (three blocks) with variable spacing
and the check that the use of an additional lens for the fixed spacing also works
we leave as an exercise for the interested reader.

In analogy with (\ref{am1_0}) the
combination of three blocks can be written as 

\noindent
\begin{eqnarray}
W^{x, y}_3 \, W^{x, y}_2 \, W^{x, y}_1 \,=\,
\nonumber
\end{eqnarray}

\noindent
\begin{eqnarray}
-\Lambda^{-1} \left(a^{x,y}\right)
P \left(\hat{w}^{x,y}_3\right)
P \left(\hat{w}^{x,y}_2\right)
P \left(\hat{w}^{x,y}_1\right)
\Lambda \left(a^{x,y}\right)
\label{am1}
\end{eqnarray}

\noindent
where

\noindent
\begin{eqnarray}
a^{x,y} \,=\, \sqrt{\frac{u^{x,y}_1 u^{x,y}_3}{u^{x,y}_2}}
\label{am2}
\end{eqnarray}

\noindent
and

\noindent
\begin{eqnarray}
\hat{w}^{x,y}_1 = \frac{u^{x,y}_2}{u^{x,y}_3}\cdot w^{x,y}_1,
\label{am3}
\end{eqnarray}

\noindent
\begin{eqnarray}
\hat{w}^{x,y}_2 = \frac{u^{x,y}_3}{u^{x,y}_1}\cdot w^{x,y}_2,
\label{am4}
\end{eqnarray}

\noindent
\begin{eqnarray}
\hat{w}^{x,y}_3 = \frac{u^{x,y}_1}{u^{x,y}_2}\cdot w^{x,y}_3.
\label{am5}
\end{eqnarray}

\noindent
Plugging again this representation into system (\ref{es1}) we obtain
the equation

\noindent
\begin{eqnarray}
P \left(\hat{w}^{x,y}_3\right)
P \left(\hat{w}^{x,y}_2\right)
P \left(\hat{w}^{x,y}_1\right) =
\nonumber
\end{eqnarray}

\noindent
\begin{eqnarray}
-\Lambda \left(a^{x,y}\right)
\hat{M}_{x,y}
\Lambda^{-1} \left(a^{x,y}\right).
\label{am1_0011}
\end{eqnarray}

\noindent
We know that the sufficient condition for this equation to be solvable with respect to
the unknowns $\hat{w}^{x,y}_k$ is
that the horizontal and vertical parts of the matrix
on the right-hand side both have nonvanishing $r_{22}$ elements. 
The direct calculation gives us

\noindent
\begin{eqnarray}
r_{22}^{x,y} \,=\,
\frac{
m_{12}^{x,y} -
m \,m_{11}^{x,y} -
p \,m_{22}^{x,y} +
m p \,m_{21}^{x,y}}{m + p},
\label{am1_0012}
\end{eqnarray}

\noindent
where $m_{ij}^{x,y}$ are the elements of the input matrix $M$.

Looking for a solution
one can proceed further in the same manner as in the four block case 
with only one difference. At the first step one has to take not arbitrary
nonnegative $m$ and $p$, but such $m$ and $p$ that both $r_{22}^x$
and $r_{22}^y$ are nonzero, which due to symplecticity of the matrices $M_x$ and $M_y$
is always possible.

\subsection{Recipe of construction of lens blocks with 
decoupled transverse actions}

In this subsection we give the recipe for the construction of lens blocks with 
decoupled transverse actions. As we will see, this recipe works not only for 
the four-lens combination considered above, but is also applicable to blocks 
with a larger number of lenses.

Let us consider $q-$lens block with $q \geq 4$:

\noindent
\begin{eqnarray}
W^{x,y} = P(2 \pm l g_q) \cdot \ldots \cdot P(2 \pm l g_1),
\label{db_0}
\end{eqnarray}

\noindent
and let us assume that the product of the $(q-2)$ inner matrices in our block takes 
the form

\noindent
\begin{eqnarray}
P(2 \pm l g_{q-1}) \cdot \ldots \cdot P(2 \pm l g_2) = 
\left(
\begin{array}{cc}
 0            & u^{x,y}\\
-1 / u^{x, y} & \tau^{x,y}
\end{array}
\right).
\label{db_1}
\end{eqnarray}

\noindent
Then, as one can show by direct multiplication, 
both matrices $W^x$ and $W^y$
become similar to a single P matrix
(with an inessential minus sign possibly presented), namely

\noindent
\begin{eqnarray}
W^{x,y} \,=\,-\mbox{sign}(u^{x,y}) \cdot
\nonumber
\end{eqnarray}

\noindent
\begin{eqnarray}
\Lambda^{-1} \left(\sqrt{\left|u^{x,y}\right|}\right)
P (w^{x,y}) \Lambda \left(\sqrt{\left|u^{x,y}\right|}\right),
\label{db_2}
\end{eqnarray}

\noindent
where

\noindent
\begin{eqnarray}
w^{x, y} = 
\frac{2 \pm l g_1 }{\left|u^{x,y}\right|}
+ \left|u^{x,y}\right| \, \left(2 \pm l g_q \right) 
+ \mbox{sign}(u^{x,y})\, \tau^{x,y}.
\label{db_3}
\end{eqnarray}

\noindent
If for arbitrary given values of $w^x$ and $w^y$ Eq. (\ref{db_3}) 
can be solved with respect to the variables 
$g_1$ and $g_q$, then it will be exactly what we need,
and the necessary and sufficient condition for such solvability is

\noindent
\begin{eqnarray}
\left|u^x\right| \;\neq \;\left|u^y\right|.
\label{db_4}
\end{eqnarray}

\noindent
So, in order to construct the $q$-lens block with the decoupled transverse actions,
one has to solve two equations making the $r_{11}$ elements of the $x$ and $y$
parts of the product of the $(q-2)$ inner matrices equal to zero and 
one has to satisfy one 
additional inequality constraint (\ref{db_4}). 

The solution for the four-lens block was already given above and is unique 
up to a sign change ($\delta = \pm 1$). Let us now consider the more complicated 
(but still analytically solvable) case of five lenses. In this situation all 
possible solutions which bring the product of the three 
inner $P$ matrices 

\noindent
\begin{eqnarray}
P(2 \pm l g_4)\, P(2 \pm l g_3)\, P(2 \pm l g_2),
\label{db_0008}
\end{eqnarray}

\noindent
to the form (\ref{db_1}) can be expressed as a function of
parameters $l$ and $g_3$ as follows: 

\noindent
\begin{eqnarray}
g_2 \,=\, \frac{1}{l} \cdot \frac{l g_3 + \delta 
\sqrt{\left((l g_3)^2 - 2\right) \cdot\left((2 l g_3)^2 - 9\right)}}
{(l g_3)^ 2 - 3},
\label{db_6}
\end{eqnarray}

\noindent
\begin{eqnarray}
g_4 \,=\, \frac{1}{l} \cdot \frac{l g_3 - \delta 
\sqrt{\left((l g_3)^2 - 2\right) \cdot\left((2 l g_3)^2 - 9\right)}}
{(l g_3)^ 2 - 3},
\label{db_7}
\end{eqnarray}

\noindent
$\delta = \pm 1$, and $l > 0$ and $g_3$ are such that

\noindent
\begin{eqnarray}
l g_3 \,\in\, \left(-\infty, \,-\sqrt{3}\right) 
\cup \left(-\sqrt{3}, \,-1.5\right] 
\cup 
\nonumber
\end{eqnarray}

\noindent
\begin{eqnarray}
\left[-\sqrt{2}, \,\sqrt{2}\right] 
\cup \left[1.5,\, \sqrt{3}\right) 
\cup \left(\sqrt{3},\, +\infty\right).
\label{db_5}
\end{eqnarray}

\noindent
To complete the block construction we have to select from all these solutions 
a subset on which the functions

\noindent
\begin{eqnarray}
u^{x,y} \,=\, 1 \,-\, (l g_2 \,\mp \,2) \cdot (l g_3 \,+\, l g_4) 
\label{db_8}
\end{eqnarray}

\noindent
satisfy the inequality (\ref{db_4}). As one can check, this can be achieved
simply by removing from the set (\ref{db_5}) the endpoints of the given set intervals, 
i.e., by removing the points $\pm 1.5$ and $\pm \sqrt{2}$. 
So we see that there are many solutions which allow us to construct from five lenses 
the block with decoupled transverse 
actions and for selecting one of them some additional optimization criteria could be involved.

Note that in the blocks constructed according to our recipe the setting of the internal 
lenses does not depend on the setting of the first and the last lenses and depends only
on the geometrical block parameters (distances between the lenses), which will be seen more 
clearly in the following section where we will consider the case of arbitrarily spaced 
thin lenses.

Note also that the horizontal and the vertical matrices between the first and 
the last lenses in the block, 
when calculated using not the $P$ matrix notation, but the original 
variables in which Eq. (\ref{a1_0}) is written

\noindent
\begin{eqnarray}
D(m) B(m, \pm g_{q-1}, p) \cdot \ldots \cdot
B(m,\pm g_{2},p) D(p) = 
\nonumber
\end{eqnarray}

\noindent
\begin{eqnarray}
D(m) S^{-1}(m, p) P(2\pm l g_{q-1}) \cdot \ldots \cdot
P(2\pm l g_{2}) \cdot
\nonumber
\end{eqnarray}

\noindent
\begin{eqnarray}
S (m, p) D(p) = 
\nonumber
\end{eqnarray}

\noindent
\begin{eqnarray}
-
\left(
\begin{array}{cc}
u^{x,y} & 0\\
1 / u^{x, y} 
+\left(u^{x, y} + \tau^{x, y}\right) / l
& 
\;\;1 / u^{x, y}
\end{array}
\right),
\label{db_9}
\end{eqnarray}

\noindent
both have $r_{12}$ elements 
equal to zero (i.e. the phase advances between the first and the last lenses in 
the block are always multiples of $180^{\circ}$),
but this alone without the inequality (\ref{db_4}) satisfied does not give us the block with 
the decoupled transverse actions.

\section{Generalization to the case of arbitrarily spaced thin lenses}

When the distances between the lenses are not equal to each other, we immediately 
lose the advantage of the cancellation of $S$ matrices between the $P$ matrices after 
substitution of the representation (\ref{a2}) into Eq. (\ref{a1_0}). 
Nevertheless, as we will show below, this case can also be treated with the tools
developed in the previous section.

Let us denote by $d_{k_1, k_2}$ the distance between the lenses with the indices
$k_1$ and $k_2$ ($k_1 \leq k_2$).
We start from the observation that for $k = 2, \ldots,n$ the following
identity holds:

\noindent
\begin{eqnarray}
S(m_k,\, p_k)\,S^{-1}(m_{k-1},\, p_{k-1}) \,=
\nonumber
\end{eqnarray} 

\noindent
\begin{eqnarray}
L\left(\frac{l_k}{d_{k-1,k}} - 1 \right)
\Lambda \left(\frac{d_{k-1,k}}{\sqrt{l_{k-1}\,l_k }} \right)
U\left(1- \frac{l_{k-1}}{d_{k-1, k}}\right),
\label{b1}
\end{eqnarray}

\noindent
which can be shown by direct multiplication and which requires
that all $l_k$ and $d_{k-1, k}$ are positive.
Note that in this identity $L$ and $U$ are the lower and upper triangular matrices
with unit diagonal elements (see Appendix A for more details).

Let us now substitute the representation (\ref{a2}) into Eq. (\ref{a1_0})
and then plug in the corresponding places the right-hand side of the 
identity (\ref{b1}).
After that the property (\ref{e11}) allows us to eliminate from the result
all $L$ and $U$ matrices  while shifting their arguments to the arguments of the
neighboring $P$ matrices, and leaving us with a
product consisting of alternating $P$ and $\Lambda$ matrices. 
Although the $\Lambda$ matrices cannot be eliminated completely, they
can be moved either on the left or on the right-hand side of all
$P$ matrices with the help of the property (\ref{e9}). 
As the last step we transfer all matrices from the left and right sides
of the obtained solid block of the $P$ matrices to the right-hand side of our equation,
hide them in the matrix $\tilde{M}_{x, y}$ and end up with the equation 

\noindent
\begin{eqnarray}
P(\tilde{v}^{x,y}_n) \cdot \ldots \cdot P(\tilde{v}^{x, y}_1) \,=\, \tilde{M}_{x, y},
\label{b2}
\end{eqnarray}

\noindent
which already has the desired form.
The detailed structure of the arguments $\tilde{v}^{x, y}_k$ and of the matrix
$\tilde{M}_{x, y}$ depends on the particular ways how the individual $\Lambda$
matrices were moved (to the left or to the right sides) and is given below 
for the case when during transformations all $\Lambda$ matrices were moved 
to the left-hand side of the $P$ matrix block.
Nevertheless, the expressions given below are general in the sense that they
contain an arbitrary positive parameter $c_1$, and with the proper choice of
this parameter one can account for all possible ways of movement of the individual
$\Lambda$ matrices:

\noindent
\begin{eqnarray}
\tilde{M}_{x,y} = \Lambda(c_n) \,S(m_n, p_n)\, M_{x, y}  S^{-1}(m_1, p_1) \Lambda(c_1),
\label{b3}
\end{eqnarray}

\noindent
\begin{eqnarray}
\tilde{v}^{x,y}_k = c_k^2 l_k \left(\frac{d_{k-1,k+1}}{d_{k-1,k} \,d_{k,k+1}} \pm g_k\right),
\;\;k = 1, \ldots, n,
\label{b4}
\end{eqnarray}

\noindent
\begin{eqnarray}
c_k \,=\, \frac{d_{k-1, k}}{\sqrt{ l_{k-1}\,l_k}} \cdot \frac{1}{c_{k-1}},
\;\;\;\;k = 2, \ldots, n,
\label{b5}
\end{eqnarray}

\noindent
$c_1$ is an arbitrary positive parameter
and, because we do not have lenses with indices $0$ and $n+1$,
we use the conventions that

\noindent
\begin{eqnarray}
d_{0,1} \,=\, l_1, 
\;\;\;\;d_{0,2} \,=\, d_{0,1} + d_{1,2},
\label{b6_1}
\end{eqnarray}

\noindent
\begin{eqnarray}
d_{n,n+1} \,=\, l_n,\;\;\;\;
d_{n-1,n+1} \,=\,d_{n-1,n} + d_{n,n+1}. 
\label{b6_2}
\end{eqnarray}

\noindent
Note that, if the parameter $c_1$ is taken to be a positive number or
a dimensionless function of the thin-lens multiplet parameters (drift lengths
and lens strengths), then Eq. (\ref{b2}) and the variables
(\ref{b4}) are also dimensionless. One of the possible choices
is to take $c_1$ for even $n$ as solution of the equation $c_n = c_1$  
and for odd $n$ as solution of the equation $c_n = c_1^{-1}$.
If the condition (\ref{es0}) holds, then the solution of these equations for 
both cases (even and odd $n$) is $c_1 = 1$ and the representation (\ref{b2}) 
turns into the representation (\ref{es1}) as one can expect.

Now in order to continue we need a lens block with the decoupled transverse actions
and, as it is not difficult to check, the recipe given in the previous section
is applicable without any changes. For the construction of the $q$-lens block we still
need to bring the product of the $(q-2)$ inner matrices to the form (\ref{db_1}) 
while also satisfying the inequality constraint (\ref{db_4}). 
For the four-lens case

\noindent
\begin{eqnarray}
W^{x,y}\,=\,P(\tilde{v}^{x,y}_{4}) \,P(\tilde{v}^{x,y}_{3}) \,
P(\tilde{v}^{x,y}_{2}) \,P(\tilde{v}^{x,y}_{1})
\label{db1}
\end{eqnarray}

\noindent
the two equations making the $r_{11}$ elements of the $x$ and $y$
parts of the product of the two inner matrices equal to zero are

\noindent
\begin{eqnarray}
\tilde{v}^{x,y}_2 \cdot \tilde{v}^{x,y}_3 = 1,
\label{db4}
\end{eqnarray}

\noindent
and have a solution 

\noindent
\begin{eqnarray}
g_{2} = \;\,\frac{\delta}{d_{1,2}} \cdot
\sqrt{
\frac{d_{1,4}}{d_{2,3}} \cdot
\frac{d_{1,3}}{d_{2,4}}},
\label{db2}
\end{eqnarray}

\noindent
\begin{eqnarray}
g_{3} = -\frac{\delta}{d_{3,4}}\cdot
\sqrt{
\frac{d_{1,4}}{d_{2,3}}\cdot
\frac{d_{2,4}}{d_{1,3}}},
\label{db3}
\end{eqnarray}

\noindent
which again is unique up to a sign change ($\delta = \pm 1$).
The values $u^{x, y}$ for this solution are 

\noindent
\begin{eqnarray}
u^{x, y} \,=\,\tilde{v}_3^{x,y}\,=\,
\frac{c_3^2 \,l_3}{d_{3, 4}} \cdot
\left(\frac{d_{2,4}}{d_{2,3}} \,\mp\, \delta \cdot
\sqrt{
\frac{d_{1,4}}{d_{2,3}}\cdot
\frac{d_{2,4}}{d_{1,3}}}
\right).
\label{db3_0}
\end{eqnarray}

\noindent
Both of them are positive and clearly satisfy the inequality (\ref{db_4}).
With this choice for $g_2$ and $g_3$ the total block matrix takes
the form

\noindent
\begin{eqnarray}
W^{x, y} \,=\, -\Lambda^{-1} \left(\sqrt{u^{x,y}}\right)\, 
P (w^{x,y})\,\Lambda \left(\sqrt{u^{x,y}}\right)
\label{db5}
\end{eqnarray}

\noindent
where

\noindent
\begin{eqnarray}
w^{x,y} \;=\;\left(u^{x,y}\right)^{-1}\cdot \tilde{v}^{x,y}_1 
\;+\; u^{x,y} \cdot \tilde{v}^{x,y}_4 \;-\;1. 
\label{db6}
\end{eqnarray}

\noindent
Equation (\ref{db6}) is the analogy of the 
formulas (\ref{es7}) and (\ref{es8}) and 
for any given values $w^x$ and $w^y$ allow one to
determine the corresponding lens strengths $g_1$ and $g_4$.
Thus, all results of the previous section 
concerning the reduction of the 2D problem to two 1D
problems become applicable with some minor changes
connected with the difference in the matrices $\hat{M}_{x,y}$
and $\tilde{M}_{x,y}$ defined by the relations (\ref{es2}) and (\ref{b3}), respectively. 
Note that if, when placed in the beam line, the actual decoupling block 
starts from the lens with the index $k$,
one has simply to add $k-1$ to the indices $1,2,3$ and $4$ in all
above formulas.

\subsection{Removing of superfluous parameters}

Equation (\ref{a1_0})
contains $2 n$ parameters which specify the drift lengths 
($m_1, p_1, \ldots , m_n, p_n$)
while only $n + 1$ parameters,
namely $m_1, d_{1,2}, \ldots, d_{n-1, n}, p_n$ have a clear physical
meaning and are independent. 
Let us have a closer look at formulas (\ref{b2})-(\ref{b6_2})
and count how many superfluous parameters are still left in them
and then show ways to remove them.

The superfluous parameters $p_1$ and $m_n$ are clearly present,
either directly as the arguments of $S$ matrices or
through the lengths of the first and the last building blocks $l_1$ and $l_n$.
And actually that is all.
The presence of the other superfluous parameters through
the values $l_2, \ldots, l_{n-1}$ is completely imaginary. 
To show this let us note that these values can enter the main formulas
(\ref{b2})-(\ref{b4}) only through the values $c_1$ and $c_n$ and
through the combinations $c_1^2 l_1, \ldots, c_n^2 l_n$. So if we
choose $c_1$ to be independent from $l_2, \ldots, l_{n-1}$, then
these parameters can enter in none of the combinations $c_k^2 l_k$ 
due to the recursion relation

\noindent
\begin{eqnarray}
c_k^2\,l_k \,=\,d_{k-1,k}^2 \cdot
\frac{1}{c_{k-1}^2\,l_{k-1}}
\;\;\;\;k = 2, \ldots , n,
\label{db8}
\end{eqnarray}

\noindent
which follows from the recursion relation (\ref{b5}), 
and likewise they cannot enter the value $c_n$ because one
can write that $c_n = \sqrt{c_n^2 l_n / l_n}$.

Thus, there are only two superfluous parameters, $p_1$ and $m_n$, 
present in our formulas, either directly or through the values $l_1$ and $l_n$.
Do we need to remove them? In general not, because it is clear that none
of the physically meaningful answers will depend on them and, in this sense,
their absence in the final results (like in formulas (\ref{db2}) and (\ref{db3})) 
could work as some indirect indicator of the correctness of the calculations.
But from another point of view, it seems better not to have
any superfluous parameters from which one can expect nothing 
except some possible additional complications. 

The simplest way to remove the parameters $p_1$ and $m_n$ from 
the formulas (\ref{b2})-(\ref{b4})
is to make them functions of the physically meaningful parameters. 
For example, one can take $p_1 = 0.5 \cdot d_{1, 2}$ and $m_n = 0.5 \cdot d_{n-1, n}$.
However, the way which we prefer is the modification of 
the formulas (\ref{b2})-(\ref{b4}) in such a way that the
superfluous parameters will disappear automatically.
In doing so let us first present the final result and then
make some remarks on how it can be obtained:

\noindent
\begin{eqnarray}
P(v^{x,y}_n) \cdot \ldots \cdot P(v^{x, y}_1) \,=\, \breve{M}_{x, y},
\label{ff_1}
\end{eqnarray}

\noindent
\begin{eqnarray}
\breve{M}_{x,y} \,=\, J\Lambda^{-1}(b_n) U(-p_n)\, M_{x, y}\,U(-m_1) \Lambda(b_1),
\label{ff_2}
\end{eqnarray}

\noindent
\begin{eqnarray}
v^{x,y}_1 = b_1^2 \left(\frac{1}{d_{1,2}} \pm g_1\right),
\label{ff_3}
\end{eqnarray}

\noindent
\begin{eqnarray}
v^{x,y}_k = b_k^2 \left(\frac{d_{k-1,k+1}}{d_{k-1,k} \,d_{k,k+1}} \pm g_k\right),
\;\,k = 2, \ldots, n-1,
\label{ff_4}
\end{eqnarray}

\noindent
\begin{eqnarray}
v^{x,y}_n = b_n^2 \left(\frac{1}{d_{n-1,n}} \pm g_n\right),
\label{ff_5}
\end{eqnarray}

\noindent
\begin{eqnarray}
b_1 > 0, \;\;\;b_k \,=\, d_{k-1, k} \cdot \frac{1}{b_{k-1}},
\;\;\;\;\;k = 2, \ldots, n,
\label{ff_6}
\end{eqnarray}

\noindent
and $J$ is the $2 \times 2$ symplectic unit matrix.

In order to obtain formulas (\ref{ff_1})-(\ref{ff_6}) from 
formulas (\ref{b2})-(\ref{b6_2})
let us first introduce the parameters $b_k = c_k \sqrt{l_k}$ and
then assume that $c_1$ is chosen in such a way that
$b_1$ does not depend on any superfluous parameter
(for example, one simply can take $c_1 = 1 / \sqrt{l_1}$).
After this one sees that the parameters $l_1$ and $l_n$ enter the left-hand side
of Eq. (\ref{b2}) only through the matrices $P(\tilde{v}^{x,y}_1)$ 
and $P(\tilde{v}^{x,y}_n)$.
Because of the property (\ref{e11}) these matrices can be 
decomposed into the following products:

\noindent
\begin{eqnarray}
P(\tilde{v}^{x,y}_1) \,=\, P(v^{x,y}_1) \, L(c_1^2)
\,=\, P(v^{x,y}_1) \, L(b_1^2 \,/\, l_1),
\label{ff_7}
\end{eqnarray}

\noindent
\begin{eqnarray}
P(\tilde{v}^{x,y}_n) = U(-c_n^2) \,P(v^{x,y}_n) 
= U(-b_n^2 \,/\, l_n)\,P(v^{x,y}_n).
\label{ff_8}
\end{eqnarray}

\noindent
As the last step, one has to substitute these decompositions 
back into Eq. (\ref{b2}),
transfer $U$ and $L$ to the right-hand side and, after
some straightforward manipulations, arrive at the final result
described in the above formulas (\ref{ff_1})-(\ref{ff_6}). 

Note that the whole story about the presence of the superfluous parameters
is the result of our desire to have the expressions for the problem 
description (expressions (\ref{b2})-(\ref{b6_2})) which reduces to  
the highly symmetric expressions (\ref{es1}) and (\ref{es2})
in the limit of equal distances between thin lenses. If one does not
require that, then, as we will outline below, it is possible to arrive
at the representation (\ref{ff_1})-(\ref{ff_6}) without using the
identity (\ref{a2}). 

According to (\ref{e5}) and (\ref{e6}) the matrix of 
the building block can be written as

\noindent
\begin{eqnarray}
B(m, \,\pm g, \,p) \,=\,P(-p) \,P(\pm g)\, P(-m)\, J.
\label{rsp_1}
\end{eqnarray}

\noindent
Substituting this representation in the original Eq. (\ref{a1_0})
and using that due to (\ref{e4_2})

\noindent
\begin{eqnarray}
P(-m_k)\,J\,P(-p_{k-1})\,=\,-P(-d_{k-1, k})
\label{rsp_2}
\end{eqnarray}

\noindent 
we obtain

\noindent
\begin{eqnarray}
P(\pm g_n) P(-d_{n-1, n}) \cdot \ldots \cdot P(-d_{1, 2}) P(\pm g_1) \Lambda(b_1) =
\nonumber
\end{eqnarray}

\noindent
\begin{eqnarray}
(-1)^{n-1} J\, U(-p_n)\,
M_{x,y} \,U(-m_1) \,\Lambda(b_1),
\label{rsp_3}
\end{eqnarray}

\noindent
where we have already introduced an arbitrary
positive parameter $b_1$.
Now, assuming that all distances between lenses are positive
and using (\ref{e4_8}), we can replace for each $k = 2, \ldots, n$ the matrix
$P(-d_{k-1, k})$  by the matrix $-\Lambda(d_{k-1, k})$
with simultaneous adding to the arguments of the two neighboring P
matrices
the value $d_{k-1, k}^{-1}$.
After these manipulations we arrive at the expression

\noindent
\begin{eqnarray}
P(d_{n-1, n}^{-1} \pm g_n)\, \Lambda(d_{n-1, n}) \cdot
\nonumber
\end{eqnarray}

\noindent
\begin{eqnarray} 
P(d_{n-1, n}^{-1} + d_{n-2, n-1}^{-1} \pm g_{n-1})\,
\Lambda(d_{n-2, n-1})
\cdot
\ldots
\nonumber
\end{eqnarray}

\noindent
\begin{eqnarray}
\ldots
\cdot
P(d_{2, 3}^{-1} + d_{1, 2}^{-1} \pm g_2)
\Lambda(d_{1, 2}) P(d_{1, 2}^{-1} \pm g_1) \Lambda(b_1) =
\nonumber
\end{eqnarray}

\noindent
\begin{eqnarray}
J\, U(-p_n)\,
M_{x,y} \,U(-m_1) \,\Lambda(b_1),
\label{rsp_4}
\end{eqnarray}

\noindent
and the last step, which is still necessary in order 
to obtain formulas (\ref{ff_1})-(\ref{ff_6}),
is to move all $\Lambda$ matrices to the left
in the left-hand side of Eq. (\ref{rsp_4})
using the identity (\ref{e9}) with a subsequent transfer of
the matrix 
$\Lambda(b_n^{-1})$ 
from the left to the right-hand side of the
obtained equality.

\begin{acknowledgments}
The authors are thankful to Winfried Decking, Nina Golubeva and Helmut Mais
for support and their interest in this work.
The careful reading of the manuscript by Helmut Mais
and his useful advices are gratefully acknowledged.
\end{acknowledgments}

\appendix

\section{Elementary matrices and their properties}

The elementary symplectic $P$ matrix which is defined as follows,

\noindent
\begin{eqnarray}
P(a) \,=\, 
\left(
\begin{array}{rr}
 a &  1 \\
-1 &  0      
\end{array}
\right)
\label{e1}
\end{eqnarray}

\noindent
and which we use extensively throughout this paper
was found empirically by the usual trial and error method
during attempts to reduce the problem of analytical
study of thin-lens multiplets to some
``more manageable'' form.
As we will see below, this matrix possesses many interesting properties
not only by itself, but also in combination with the other elementary
matrices. 
Although not widely known in the scientific community,  
it was no surprise, as we found later, that it 
was successfully 
used in some special area of abstract algebra \cite{Cohn}.

In order to give an expression for the product of 
$n$ elementary $P$ matrices, let us first define  
a sequence of polynomials $\kappa_n$ in the variables
$z_1, \ldots , z_n$ recursively by
the following equations:

\noindent
\begin{eqnarray}
\kappa_{-1}\,=\,0,
\;\;\;\;\;
\kappa_{0}\,=\,1,
\label{e1_r1}
\end{eqnarray}

\noindent
\begin{eqnarray}
\kappa_n (z_1, \ldots , z_n) \,=\, z_n \cdot \kappa_{n-1} (z_1, \ldots , z_{n-1})\,- 
\nonumber
\end{eqnarray}

\noindent
\begin{eqnarray}
\kappa_{n-2} (z_1, \ldots , z_{n-2}),
\;\;\;\;\; n \,\geq\, 1.
\label{e1_r2}
\end{eqnarray}

\noindent
With these notations we assert that

\noindent
\begin{eqnarray}
P(a_n)\cdot \ldots \cdot P(a_1) \,=
\nonumber
\end{eqnarray}

\noindent
\begin{eqnarray}
\left(
\begin{array}{cc}
 \kappa_n (a_1, \ldots , a_n) & \kappa_{n-1} (a_2, \ldots , a_n)\\
-\kappa_{n-1} (a_1, \ldots , a_{n-1}) & -\kappa_{n-2} (a_2, \ldots , a_{n-1})
\end{array}
\right),
\label{e1_r3}
\end{eqnarray}

\noindent
which is clear for $n = 1$ and in the general case can be proven by induction.
Because such induction can be made in two different ways, either by adding
one more $P$ matrix from the left or from the right side, it is easy to see
that the polynomials $\kappa_n$ can also be defined by (\ref{e1_r1}) and
by the recursion relation

\noindent
\begin{eqnarray}
\kappa_n (z_1, \ldots , z_n) \,=\, z_1 \cdot \kappa_{n-1} (z_2, \ldots , z_n)\,- 
\nonumber
\end{eqnarray}

\noindent
\begin{eqnarray}
\kappa_{n-2} (z_3, \ldots , z_n),
\;\;\;\;\; n \,\geq\, 1.
\label{e1_r4}
\end{eqnarray}

\noindent
Comparison of (\ref{e1_r2}) and (\ref{e1_r4}) implies that

\noindent
\begin{eqnarray}
\kappa_n (z_1, z_2, \ldots,z_{n-1}, z_n) \equiv  
\kappa_n (z_n, z_{n-1}, \ldots, z_2, z_1).
\label{e1_r5}
\end{eqnarray}

\noindent
According to (\ref{e1_r3}) we can write down the matrix of the product
of any number of elementary $P$ matrices without making any matrix multiplications.
In this connection let us note that the problem of deriving some
recursion relations which allow one to obtain the 
transfer matrix of an arbitrary multiplet without
actual matrix multiplications was also addressed in \cite{Regenstreif_1}.

It is clear that the matrix $P(0)$ coincides with the $2 \times 2$ 
symplectic unit matrix $J$, i.e., that

\noindent
\begin{eqnarray}
P(0) \,=\, 
\left(
\begin{array}{rr}
 0 &  1 \\
-1 &  0      
\end{array}
\right) \,=\, J,
\label{e4}
\end{eqnarray}

\noindent
and the following relations between the $P$ matrices
can be easily verified by direct multiplication:

\noindent
\begin{eqnarray}
P(a) \,J\, P(b)\,=\, -P(a + b),  
\label{e4_2}
\end{eqnarray}

\noindent
\begin{eqnarray}
P^3(\pm 1) \,=\,\mp I,   
\label{e4_4}
\end{eqnarray}

\noindent
\begin{eqnarray}
P^{-1}(a) \,=\, J \,P(-a)\, J   
\,=\,a \cdot I \,-\,P(a),
\label{e4_5}
\end{eqnarray}

\noindent
\begin{eqnarray}
P(a) \,P^{-1}(b) \,=\, - P(a-b)\, J,   
\label{e4_6}
\end{eqnarray}

\noindent
\begin{eqnarray}
P(a) \,P^{-1}(b) \,P(c) \,=\, P(a-b+c),
\label{e4_7}
\end{eqnarray}

\noindent
where $I$ is the $2 \times 2$ identity matrix.

Let us now introduce three more
elementary matrices. The diagonal (scaling) matrix

\noindent
\begin{eqnarray}
\Lambda(a) \,=\, 
\left(
\begin{array}{cc}
a &  0 \\
0 &  a^{-1}      
\end{array}
\right)
\label{e8}
\end{eqnarray}

\noindent
and the lower and upper triangular matrices with unit diagonal
elements

\noindent
\begin{eqnarray}
L(a) \,=\, 
\left(
\begin{array}{rr}
1 &  0 \\
a &  1      
\end{array}
\right),
\;\;\;\;\;
U(a) \,=\, 
\left(
\begin{array}{rr}
1 &  a \\
0 &  1      
\end{array}
\right).
\label{e3}
\end{eqnarray}

\noindent
Note that although the matrices $L$ and $U$ formally coincide with the matrices
of the thin lens and the drift space, respectively, we have introduced them
in order to distinguish the situations where matrix of 
lens or drift
has physical meaning and where the usage of low or upper triangular
matrix is simply the reflection of the mathematical technique used.

We have the following relations between the matrices
$P$, $\Lambda$, $L$ and $U$:

\noindent
\begin{eqnarray}
P(a) \,P(a^{-1})\, P(a)\,=\,\Lambda(-a),
\label{e4_3}
\end{eqnarray}

\noindent
\begin{eqnarray}
P(a) \,P(b^{-1}) \,P(c) \,=\, P(a-b) \,\Lambda(b^{-1}) \,P(c-b),   
\label{e4_8}
\end{eqnarray}

\noindent
\begin{eqnarray}
\Lambda(a)\,P(b) \,\Lambda(a) \,=\,P(a^2 b),
\label{e9}
\end{eqnarray}

\noindent
\begin{eqnarray}
P(a^{-1})\,=\, L(-a)\, \Lambda(a^{-1})\, U(a),
\label{e4_1}
\end{eqnarray}

\noindent
\begin{eqnarray}
U(a)\,P(b) \,L(c) \,=\,P(b + c - a), 
\label{e11}
\end{eqnarray}

\noindent
\begin{eqnarray}
L(a) \,=\, -J \,P(a),
\label{e5}
\end{eqnarray}

\noindent
\begin{eqnarray}
U(a) \,=\, -P(-a)\,J.
\label{e6}
\end{eqnarray}

\noindent
Although these relations are elementary, they
are basic for all results of this paper.

\section{Three Explicit Solutions for Four-Lens Telescopes}

A telescope is a beam transport system which has diagonal transfer matrices
in both transverse planes,

\noindent
\begin{eqnarray}
M_{x,y}\,=\,
\left(
\begin{array}{cc}
\mbox{\ae}_{x,y} & 0\\
0 & \mbox{\ae}_{x,y}^{-1}
\end{array}
\right),
\label{FLT_01}
\end{eqnarray}

\noindent
where the numbers $\mbox{\ae}_x$ and $\mbox{\ae}_y$
are called magnifications (or de-magnifications, if convenient;
negativity of horizontal/vertical magnification means that 
the horizontal/vertical image is inverted with respect to the original).
It is an optics module which is important for many
accelerator designs and its study has received considerable
attention in the past (see, for example, papers
~\cite{BrownServranckx,MontagueRuggiero,Zotter,Napoly,Autin}).
The minimum number of thin lenses required for a telescope to exist
is believed to be four (though, to our knowledge, still no rigorous proof is available) 
and the corresponding four-lens telescope system 
of matrix equations 
in our notations
can be written as follows:

\noindent
\begin{eqnarray}
D(p_4)\,Q(\pm g_4)\,D(d_{3,4})\,Q(\pm g_3)\,D(d_{2,3})\cdot
\nonumber
\end{eqnarray}

\noindent
\begin{eqnarray}
Q(\pm g_2)\,D(d_{1,2})\,Q(\pm g_1)\,D(m_1) \,=\, M_{x,y}.
\label{FLT_01_01}
\end{eqnarray}

There are two explicit analytical solutions known
for the system (\ref{FLT_01_01}).
The first solution is obtained when the astronomical
telescope, consisting of two focusing lenses separated by
the distance equal to the sum of their focal lengths, is generalized
to the usage of magnetic doublets instead of optical lenses.
This solution has the property that  

\noindent
\begin{eqnarray}
\mbox{\ae}_x \,=\, \mbox{\ae}_y \, < \, 0,
\label{FLT_01_014}
\end{eqnarray}

\noindent
i.e. it always provides telescopes with
equal negative magnifications in both transverse planes
(see, for example ~\cite{MontagueRuggiero}).
The second known analytical solution in the four-lens case is 
the solution for an inversor ~\cite{Autin}, 
which is the name of the telescope with
the horizontal and vertical magnifications being inverse of one another: 

\noindent
\begin{eqnarray}
\mbox{\ae}_x \cdot \mbox{\ae}_y \,=\, 1,
\;\;\;\;\;
\mbox{\ae}_{x,y} \,<\,0.
\label{FLT_01_015}
\end{eqnarray}

Besides these two explicit solutions, all other studies of the
four-lens telescopes (as well as telescopes constructed from larger
number of lenses) are either purely numerical or semianalytical
as in ~\cite{Zotter,Napoly,Autin}, where in the first step the part
of variables is eliminated from the system (\ref{FLT_01_01}) analytically
and, in the second step, the remaining equations are solved numerically.
Since these remaining equations are not linear in the variables,
they cannot be solved easily even numerically and, therefore,
any new explicit solution of the system (\ref{FLT_01_01}) is of
interest. In this Appendix we provide new
analytical solutions using tools and techniques developed in this paper. 

Let us first transform the system (\ref{FLT_01_01})
to the $P$ matrix representation (\ref{ff_1})-(\ref{ff_6}).
If we take $b_1$ in (\ref{ff_6}) as follows,

\noindent
\begin{eqnarray}
b_1\,=\,
\sqrt{\frac{d_{1,2}\, d_{3,4}}{d_{2,3}}},
\label{FLT_02}
\end{eqnarray}

\noindent
then, after some straightforward manipulations, we obtain
the equations

\noindent
\begin{eqnarray}
P(v^{x,y}_4)\,P(v^{x, y}_3)\,P(v^{x, y}_2)\,P(v^{x, y}_1) 
\,=\, \breve{M}_{x, y},
\label{FLT_07}
\end{eqnarray}

\noindent
where

\noindent
\begin{eqnarray}
\breve{M}_{x,y}\,=\,
\left(
\begin{array}{cc}
0 & \mbox{\ae}_{x,y}^{-1}\\
-\mbox{\ae}_{x,y} & \nu_{x,y}
\end{array}
\right),
\label{FLT_04}
\end{eqnarray}

\noindent
\begin{eqnarray}
\nu_{x,y}\,=\,
\frac{d_{2,3}}{d_{1,2}\,d_{3,4}}
\,(m_1 \,\mbox{\ae}_{x,y} \,+\, p_4 \,\mbox{\ae}_{x,y}^{-1}),
\label{FLT_05}
\end{eqnarray}

\noindent
and the $P$ matrix arguments $v^{x, y}_k$ are given by
the formulas (\ref{ff_3})-(\ref{ff_5}).

The equivalent to the eight equations (\ref{FLT_07})
system of the six independent equations 
was already obtained in the course of this paper.
To get it, one simply  has to substitute $v^{x, y}_k$ instead of
$z_k$ and elements of the matrix $\breve{M}_{x,y}$
instead of $m_{kl}$ into the system (\ref{s1d6}).
The resulting system can be further simplified
taking into account the special form of the matrix $\breve{M}_{x,y}$.
If $\nu_{x,y}\,\neq\,0$, then Eq. (\ref{FLT_07}) is equivalent 
to the system

\noindent
\begin{eqnarray}
\left\{
\begin{array}{l}
\vspace{0.2cm}
v^{x,y}_2 \,+\, \nu_{x,y}\, v^{x,y}_4 \,=\, -\mbox{\ae}_{x,y}^{-1}\\
\vspace{0.2cm}
v^{x,y}_3 \,+\, \nu_{x,y} \,v^{x,y}_1 \,=\, -\mbox{\ae}_{x,y}\\ 
v^{x,y}_2 \,v^{x,y}_3 \,=\, 1 \,-\,\nu_{x,y}
\end{array}
\right.
\label{FLT_08}
\end{eqnarray}

\noindent
and if $\nu_{x,y}\,=\,0$ 
($\nu_{x}$  and $\nu_{y}$ can be zero or nonzero only simultaneously),
then the equivalent system takes on the form

\noindent
\begin{eqnarray}
\left\{
\begin{array}{l}
\vspace{0.2cm}
v^{x,y}_2 \,=\, -\mbox{\ae}_{x,y}^{-1}\\
\vspace{0.2cm}
v^{x,y}_3 \,=\, -\mbox{\ae}_{x,y}\\ 
\mbox{\ae}_{x,y}^{-1}\, v^{x,y}_1 \,+\, \mbox{\ae}_{x,y}\, v^{x,y}_4 \,=\, -1
\end{array}
\right.
\label{FLT_10}
\end{eqnarray}

It is intuitively clear that the two known analytical solutions
for the four-lens telescopes are somehow connected with the
symmetry relations (\ref{FLT_01_014}) and (\ref{FLT_01_015}),
but it is not obvious, 
when looking directly at the telescope matrix (\ref{FLT_01}),
how to find other symmetry conditions, which could allow us
to find new explicit solutions.
One of the advantages of the $P$ matrix representation 
of Eq. (\ref{FLT_01_01}) is that 
the form of the matrix $\breve{M}_{x,y}$
in (\ref{FLT_07}) gives us a useful hint that as such 
a symmetry condition one may try the condition

\noindent
\begin{eqnarray}
\nu_{x}\,=\,\nu_{y}.
\label{FLT_05_01}
\end{eqnarray}

\noindent
This condition, in the next turn, can be 
considered as a combination of
the following three cases: 

\noindent
\begin{eqnarray}
m_1\,=\,p_4\,=\,0,
\label{FLT_05_03}
\end{eqnarray}

\noindent
\begin{eqnarray}
\mbox{\ae}_{x}\,=\,\mbox{\ae}_{y},
\label{FLT_05_02}
\end{eqnarray}

\noindent
\begin{eqnarray}
p_4\,=\,\mbox{\ae}_{x} \mbox{\ae}_{y}\,m_1.
\label{FLT_05_04}
\end{eqnarray}

\noindent
The condition (\ref{FLT_05_01}) is satisfied if and only if
at least one from the conditions (\ref{FLT_05_03})-(\ref{FLT_05_04})
is true.

As we will see below, all three cases 
(\ref{FLT_05_03})-(\ref{FLT_05_04})
are actually analytically solvable and, moreover, include as their parts
both previously known solutions.  
But, before giving the details,
let us make one more useful preparatory step.

As it is well known, the telescope matrix (\ref{FLT_01}) 
is invariant under a scale transformation.
It means that if the set

\noindent
\begin{eqnarray}
m_1,\, g_1,\,
d_{1,2},\, g_2,\,
d_{2, 3},\, g_3,\,
d_{3, 4},\, g_4,\, 
p_4
\label{FLT_10_01}
\end{eqnarray}

\noindent
is the solution of the system (\ref{FLT_01_01}),
then so is the set

\noindent
\begin{eqnarray}
\lambda m_1,\,     \frac{g_1}{\lambda},\,
\lambda d_{1,2},\, \frac{g_2}{\lambda},\,
\lambda d_{2,3},\, \frac{g_3}{\lambda},\,
\lambda d_{3,4},\, \frac{g_4}{\lambda},\, 
\lambda p_4,
\label{FLT_10_02}
\end{eqnarray}

\noindent
where $\lambda$ is an arbitrary positive number.
That allows us in all further considerations 
to set the length of the middle drift $d_{2,3}$
equal to one chosen unit of length 

\noindent
\begin{eqnarray}
d_{2,3}\,=\,1.  
\label{FLT_01_02}
\end{eqnarray}

\subsection{Telescopes which start and end by lens}

If the condition (\ref{FLT_05_03}) is satisfied, then
$\,\nu_{x}\,=\,\nu_{y}\,=\,0$ and 
the equivalent to the equations (\ref{FLT_07}) system is
the system (\ref{FLT_10}). 
The necessary and sufficient conditions for  this system
to have solutions are that

\noindent
\begin{eqnarray}
\mbox{\ae}_{x,y}\,<\,0
\;\;\;\;\;
\mbox{and}
\;\;\;\;\;
\mbox{\ae}_{x} \,\neq\,\mbox{\ae}_{y}.
\label{FLT_10_1}
\end{eqnarray}

\noindent
If these conditions are satisfied, then the solution
is unique (with the precision up to the scale transformation (\ref{FLT_10_02}))
and is given by the following formulas:

\noindent
\begin{eqnarray}
d_{1,2} =\frac{2 (a_1 - a_2)}{a_3^2},
\;\;\;\;
d_{3,4} =\frac{a_1 (2 - a_2)}{a_3^2},
\label{FLT_12}
\end{eqnarray}

\noindent
\begin{eqnarray}
g_1 \,=\,-\frac{a_3 \,\big(a_2^3 \,-\, 2\, a_1\, (a_1 - a_2) \,-\, 4\, a_4\big)}
{4 \,a_2 \,(2 - a_2)\,(a_1 - a_2)},
\label{FLT_13}
\end{eqnarray}

\noindent
\begin{eqnarray}
g_2 \,=\, \frac{a_3 \,(2 - a_2)}
{2 \,(a_1 - a_2)},
\;\;\;\;
g_3 \,=\, -\frac{a_3 \,(a_1 - a_2)}
{a_1 \,(2 - a_2)},
\label{FLT_14}
\end{eqnarray}

\noindent
\begin{eqnarray}
g_4 \,=\, \frac{a_3 \,\big(a_2^3 \,-\, 2\, a_1\, (2 - a_2) \,-\, 2\,a_1\, a_4\big)}
{2 \,a_1\, a_2 \,(2 - a_2)\,(a_1 - a_2)},
\label{FLT_15}
\end{eqnarray}

\noindent
where we have used the notations

\noindent
\begin{eqnarray}
\left\{
\begin{array}{lccllcc}
\vspace{0.1cm}
a_1 &=& 2 \mbox{\ae}_x \mbox{\ae}_y,
&\;\;a_2 &=& \mbox{\ae}_x + \mbox{\ae}_y,\\
a_3 &=& \mbox{\ae}_x - \mbox{\ae}_y,
&\;\;a_4 &=& \mbox{\ae}_x^2 + \mbox{\ae}_y^2.
\end{array}
\right.
\label{FLT_11}
\end{eqnarray}

\noindent
As a partial case this solution includes 
a new inversor with zero entrance and exit drifts:

\noindent
\begin{eqnarray}
\mbox{\ae}_x \,=\, \mbox{\ae}_y^{-1} \,=\, \mbox{\ae},
\label{FLT_11_01}
\end{eqnarray}

\noindent
\begin{eqnarray}
d_{1,2} \,=\, d_{3, 4} \,=\, -\frac{2\,\mbox{\ae}}{(1\,+\,\mbox{\ae})^2},
\label{FLT_11_1}
\end{eqnarray}

\noindent
\begin{eqnarray}
g_3 \,=\, -g_2 \,=\,2\,g_1\,=\,-2\,g_4\,=\, \frac{1\,-\,\mbox{\ae}^2}{2\,\mbox{\ae}}.
\label{FLT_11_2}
\end{eqnarray}

\subsection{Telescopes with equal magnifications in both transverse planes}

Now we turn our attention to the condition (\ref{FLT_05_02}) and will
consider telescopes with equal magnification in both transverse planes:

\noindent
\begin{eqnarray}
\mbox{\ae}_{x}\,=\,\mbox{\ae}_{y}\,=\,\mbox{\ae}.
\label{FLT_06_001}
\end{eqnarray}

\noindent
According to the result of the previous subsection, none of such
telescopes can exist if $m_1 = p_4 = 0$ and therefore 
the system under study is the system (\ref{FLT_08}).
By elementary analysis one can show that 
the necessary and sufficient conditions for  
the telescope with equal magnifications to exist are that

\noindent
\begin{eqnarray}
m_1^2 \,+\,p_4^2 \,>\,0
\;\;\;\;\;
\mbox{and}
\;\;\;\;\;
\mbox{\ae} \,<\, 0.
\label{FLT_06}
\end{eqnarray}

\noindent
If these conditions are satisfied, then 
all possible solutions can be expressed as follows:

\noindent
\begin{eqnarray}
d_{3,4} \,=\, |\mbox{\ae}| \,d_{1,2},
\;\;\;\;
p_4 \,=\, |\mbox{\ae}| \,(1 \,-\, |\mbox{\ae}| \,m_1),
\label{FLT_19}
\end{eqnarray}

\noindent
\begin{eqnarray}
g_1 \,=\, \delta \,
\sqrt{\,\frac{1 \,+\, |\mbox{\ae}|}{d_{1,2}}
\cdot
\frac{1 \,+\, |\mbox{\ae}| \,d_{1,2}}{1 \,+\, d_{1,2}}\,},
\label{FLT_16}
\end{eqnarray}

\noindent
\begin{eqnarray}
g_2 \,=\, -\delta \,
\sqrt{\,\frac{1 \,+\, |\mbox{\ae}|}{d_{1,2}}
\cdot
\frac{1 \,+\, d_{1,2}}{1 \,+\, |\mbox{\ae}| \,d_{1,2}}\,},
\label{FLT_17}
\end{eqnarray}

\noindent
\begin{eqnarray}
g_3 \,=\, g_1 \,/\, |\mbox{\ae}|,
\;\;\;\;\;
g_4 \,=\, g_2\,/\,|\mbox{\ae}|,
\label{FLT_18}
\end{eqnarray}

\noindent
where the free parameters are
$\,\delta = \pm 1$, $\,d_{1,2} > 0\,$ and $\,m_1\,$
satisfying the inequality

\noindent
\begin{eqnarray}
0 \,\leq\, m_1 \,\leq\, 1 \,/\, |\mbox{\ae}|.
\label{FLT_19_1}
\end{eqnarray}

Let us divide the central interval $d_{2,3} = 1$ into two parts
of the lengths $1 -|\mbox{\ae}| m_1$ and $|\mbox{\ae}| m_1$ respectively
and prescribe these subintervals to the first and to the second 
doublet cells correspondingly.
Comparing now the obtained above doublet settings with the
settings provided by the generalization of the astronomical
telescope, one can find that they coincide. 
But though this solution is already known, we 
still made a useful step. We proved that
it is the only solution for the four-lens telescope with equal magnifications
possible.

\subsection{Telescopes with nonequal magnifications and 
special ratio of entrance and exit drifts}

The remaining case to analyze is the case (\ref{FLT_05_04}),
which we will study under the additional assumptions that
the length of the entrance drift $m_1$ is nonzero and
that the horizontal and vertical magnifications are not equal
to each other, because these situations were already considered
in the previous subsections, i.e., we will study
telescopes with nonequal magnifications and with 
the special ratio of the entrance and exit drifts given by 
the relation (\ref{FLT_05_04}).
The system for analysis is the system (\ref{FLT_08}), 
and the necessary and sufficient conditions for such a telescope
to exist are that $\mbox{\ae}_{x,y} \, < \, 0$.
If these conditions are satisfied, then all possible
solutions can be expressed as follows:

\noindent
\begin{eqnarray}
d_{1,2} \,=\,\frac{2 \,(a_1 - a_2)}{a_3^2}
\cdot(1 + a_2 \,m_1),
\label{FLT_21}
\end{eqnarray}

\noindent
\begin{eqnarray}
d_{3,4} \,=\,\frac{a_1 \,(2 - a_2)}{a_3^2}
\cdot(1 + a_2 \,m_1),
\label{FLT_22}
\end{eqnarray}

\noindent
\begin{eqnarray}
p_4 \,=\, \mbox{\ae}_x \,\mbox{\ae}_y \,m_1,
\label{FLT_22_1}
\end{eqnarray}

\noindent
\begin{eqnarray}
g_2 =\frac{\delta}{d_{1,2}} 
\sqrt{\frac{1 + d_{1,2}}{1 + d_{3,4}} \cdot
(1 + a_2 m_1 + d_{1,2} + d_{3,4})},
\label{FLT_24}
\end{eqnarray}

\noindent
\begin{eqnarray}
g_3 = -\frac{\delta}{d_{3,4}} 
\sqrt{\frac{1 + d_{3,4}}{1 + d_{1,2}}\cdot
(1 + a_2 m_1 + d_{1,2} + d_{3,4})},
\label{FLT_25}
\end{eqnarray}

\noindent
\begin{eqnarray}
g_1 \,=\, -\frac{1}{a_2 \, m_1}
\cdot \left(
\frac{a_3}{2} \,+\, 
\frac{d_{3,4}}{d_{1,2}}
\cdot g_3\right),
\label{FLT_26}
\end{eqnarray}

\noindent
\begin{eqnarray}
g_4 \,=\, \frac{1}{a_2 \, m_1}
\cdot \left(
\frac{a_3}{a_1} \,-\, 
\frac{d_{1,2}}{d_{3,4}}
\cdot g_2\right),
\label{FLT_27}
\end{eqnarray}

\noindent
where the $a_k$ are given by the formulas (\ref{FLT_11})
and the free parameters are $\delta = \pm 1$ and $m_1$
satisfying the inequality

\noindent
\begin{eqnarray}
0 \,<\, m_1 \,<\, 1 \,/\,|\mbox{\ae}_x + \mbox{\ae}_y|.
\label{FLT_23}
\end{eqnarray}

The solution (\ref{FLT_21})-(\ref{FLT_27}) has
two continuous branches which are defined by the value
of the parameter $\delta$.
In the limit $m_1 \rightarrow 0$ the branch
corresponding to 

\noindent
\begin{eqnarray}
\delta \,=\, \mbox{sign}(\mbox{\ae}_x \,-\, \mbox{\ae}_y)
\label{FLT_23_001}
\end{eqnarray}

\noindent
survives and converges to the solution
(\ref{FLT_12})-(\ref{FLT_15}), and the other branch
diverges with $g_1$ and $g_4$ going to infinity.

The solution (\ref{FLT_21})-(\ref{FLT_27}) also includes
inversors described by the following formulas:

\noindent
\begin{eqnarray}
\mbox{\ae}_x \,=\, \mbox{\ae}_y^{-1} \,=\, \mbox{\ae},
\label{FLT_23_002}
\end{eqnarray}

\noindent
\begin{eqnarray}
d_{1,2} \,=\, d_{3, 4} \,=\, -2 \cdot
\frac{\mbox{\ae}\,+\,m_1\,+\,\mbox{\ae}^2\,m_1}{(1\,+\,\mbox{\ae})^2},
\label{FLT_23_1}
\end{eqnarray}

\noindent
\begin{eqnarray}
p_4 \,=\, m_1, 
\;\;\;\;\;
g_3 \,=\, -g_2, 
\;\;\;\;\;
g_4 \,=\, -g_1,
\label{FLT_23_2}
\end{eqnarray}

\noindent
\begin{eqnarray}
g_1 \,=\, \frac{1}{m_1 \,(1 \,+\, \mbox{\ae}^2)}
\left( \frac{1 \,-\,\mbox{\ae}^2}{2} \,+\,\mbox{\ae}\,g_2 \right),
\label{FLT_23_3}
\end{eqnarray}

\noindent
\begin{eqnarray}
g_2 \,=\, \delta \frac{|1\,-\,\mbox{\ae}^2|}
{2\,\sqrt{\mbox{\ae}\,(\mbox{\ae}\,+\,m_1\,+\,\mbox{\ae}^2\,m_1)}}.
\label{FLT_23_4}
\end{eqnarray}

\noindent
Note that these inversors include the previously known 
solution for the inversor ~\cite{Autin} as a partial case.
To see that one has first to set $\delta =\mbox{sign} (\mbox{\ae}^2 - 1)$
and $m_1 = 1\, /\, \lambda$, where

\noindent
\begin{eqnarray}
\lambda \,=\, \frac{(1 \,-\,\mbox{\ae})^4}{8\, \mbox{\ae}^2},
\label{FLT_23_4786}
\end{eqnarray}

\noindent
in the solution (\ref{FLT_23_002})-(\ref{FLT_23_4}),
and then scale the result with $\lambda$ 
according to the formulas (\ref{FLT_10_02}).

\section{FODO-Type Beam Line for Independent Scan of 
Horizontal and Vertical Phase Advances}

In this Appendix we will apply the explicit solution 
developed in this paper to the design of a beam line
which allows an independent scan of the horizontal and the vertical
phase advances while preserving the entrance and exit matching
conditions for the Twiss parameters.
Even though the purely numerical approach to this problem 
could result in a smaller number of lenses than we will use,
it is not an easy task. Besides the requirements to 
cover the specified range of phase advances and to preserve
the entrance and exit matching condition,  
there are a number of additional constraints which 
one has to satisfy. They could include, for example, limitations
on the lens strengths and limitations on the changes in the 
behavior of the betatron functions inside the beam line 
during the phase scan. For each lens, the outcome
of the numerical optimization is a two-dimensional array of
the lens settings corresponding to the chosen grid in the
space of phase advances.
Every change in the design specifications (which often happens
during the design stage) results in the necessity to repeat
all optimization procedures with no warranty that the new output will
be close to the previous one even for relatively
small changes in the input requirements. 

The advantage of our approach is that
most of the design problems can be addressed
without resorting to unguided numerical calculations and that
the lens settings required for obtaining the needed horizontal and
vertical phase advances can be calculated according
to explicit analytical formulas.

Note that our interest in this problem is motivated by the desire 
to have in the future the possibility for minimization of emittance growth 
due to coherent synchrotron radiation (CSR) at the European XFEL Facility ~\cite{XFEL}
by optimizing the phase advance between two bunch compressors chicanes. 

Let us consider a FODO cell of the length $L$ which begins
with a drift space of the length $L/4$ and let us assume that
the first lens is horizontally focusing with the absolute value
of its strength equal to the value

\noindent
\begin{eqnarray}
g \,=\,\frac{2 \,\sqrt{3}}{L}.
\label{FODO_1}
\end{eqnarray}

\noindent
It is a FODO cell with $120^{\circ}$ phase advance
and its periodic Twiss parameters are as follows:

\noindent
\begin{eqnarray}
\beta_{x,y} \,=\,\frac{5 \,L}{4\,\sqrt{3}},
\;\;\;\;\;
\alpha_{x,y} \,=\,\mp 2.
\label{FODO_1_1}
\end{eqnarray}

\noindent
Let us now take a string of six such FODO cells.
Comparing the value (\ref{FODO_1}) with the values
(\ref{es4}), one sees that if we freeze the
settings of the six lenses to their original
FODO settings

\noindent
\begin{eqnarray}
\left\{
\begin{array}{lcr}
\vspace{0.1cm}
g_2 = g_6 = g_{10} &=&  g\\
g_3 = g_7 = g_{11} &=& -g
\end{array}
\right.
\label{FODO_2}
\end{eqnarray}

\noindent
and allow the strengths of the remaining lenses
to be variable parameters, then we will obtain
the sequence of three four-lens blocks with
decoupled transverse actions.
We know that with the help of three blocks we
can represent most of the $4 \times 4$ uncoupled
transfer matrices and let us see what range of phase
advances our beam line can cover while preserving
the periodic matching conditions (\ref{FODO_1_1})
for the Twiss parameters.
To keep this matching 
and, in the same time, to have predefined fractional parts of 
phase advances $\mu_{x}$ and $\mu_{y}$,
the overall transfer matrix of
our beam line must have the form

\noindent
\begin{eqnarray}
M_{x,y} \,=\,
T_{x,y}^{-1} \cdot R(\mu_{x,y}) \cdot T_{x,y},
\label{Equ_04}
\end{eqnarray}

\noindent
where

\noindent
\begin{eqnarray}
T_{x,y} \,=\, 
\left(
\begin{array}{cc}
\vspace{0.1cm}
1 / \sqrt{\beta_{x,y}} & 0 \\
\alpha_{x,y} / \sqrt{\beta_{x,y}} & 
\sqrt{\beta_{x,y}}
\end{array}
\right),
\label{Equ_05}
\end{eqnarray}

\noindent
$\beta_{x,y}$ and $\alpha_{x,y}$ are the same as in (\ref{FODO_1_1}),
and $\,R(\mu_{x,y})\,$ is a $2 \times 2$ rotation matrix 

\noindent
\begin{eqnarray}
R(\mu_{x,y}) \,=\, 
\left(
\begin{array}{rr}
\vspace{0.1cm}
\cos(\mu_{x,y}) & \sin(\mu_{x,y}) \\
-\sin(\mu_{x,y}) & \cos(\mu_{x,y}) 
\end{array}
\right).
\label{Equ_05_0}
\end{eqnarray}

\noindent
Following now the procedure described in Sec. III of this paper,
one finds that for all $\mu_{x,y} \neq 240^{\circ}$ the 
matrix (\ref{Equ_04}) can be represented by three blocks
with decoupled transverse actions and that 
for all $\mu_{x,y} \neq 60^{\circ}$ the solution for
the lens strengths is unique. As concerning points
where either $\mu_{x}$ or $\mu_{y}$ is equal to 
$60^{\circ}$, there are many solutions, but it is possible 
to choose one of them such that on the whole set
$\mu_{x,y} \neq 240^{\circ}$ the lens strengths will be
continuous functions of the phase advances.
The final formulas for the lens settings can be summarized as follows:

\noindent
\begin{eqnarray}
g_1 = g_9 = -g \cdot 
\frac{28 \,-\, u^y \cdot w_1^x \,-\, u^x \cdot w_1^y}{24},
\label{FODO_4}
\end{eqnarray}

\noindent
\begin{eqnarray}
g_4 = g_{12} = g \cdot 
\frac{28 \,-\, u^x \cdot w_1^x \,-\, u^y \cdot w_1^y}{24},
\label{FODO_5}
\end{eqnarray}

\noindent
\begin{eqnarray}
g_5 =  -g \cdot 
\frac{28 \,-\, u^y \cdot w_2^x \,-\, u^x \cdot w_2^y}{24},
\label{FODO_6}
\end{eqnarray}

\noindent
\begin{eqnarray}
g_8 = g \cdot 
\frac{28 \,-\, u^x \cdot w_2^x \,-\, u^y \cdot w_2^y}{24},
\label{FODO_7}
\end{eqnarray}

\noindent
where

\noindent
\begin{eqnarray}
u^{x,y} \,=\, 2 \mp \sqrt{3}
\label{FODO_3}
\end{eqnarray}

\noindent
and

\noindent
\begin{eqnarray}
w_2^{x,y} \,=\, \cos(\mu_{x,y}) \,-\, 
\sin(\mu_{x,y}) \,/\, \sqrt{3},
\label{FODO_3_1}
\end{eqnarray}

\noindent
\begin{eqnarray}
w_1^{x,y} \,=\,w_3^{x,y} \,=\, \frac{1 \,-\, 
2\,\sin(\mu_{x,y})\,/\,\sqrt{3}}{w_2^{x,y}}
\label{FODO_3_2}
\end{eqnarray}

\noindent
for $\mu_{x,y} \neq 60^{\circ}, 240^{\circ}$, and

\noindent
\begin{eqnarray}
w_2^{x,y} \,=\,0,
\;\;\;\;
w_1^{x,y} \,=\,w_3^{x,y} \,=\, 1 / 2
\label{FODO_3_3}
\end{eqnarray}

\noindent
for $\mu_{x,y} = 60^{\circ}$.

What is in particular interesting in this solution
is the fact that, though with changing phase advances
the setting of six lenses varies, only four
independent tuning knobs are required ($w_1^{x,y}$ and $w_2^{x,y}$).

If $\mu_x$ and/or $\mu_y$ approach the value $240^{\circ}$, then
the strengths of some lenses in our solution go to infinity, but
if we restrict the region of our interest, for example, to the
region $0^{\circ} \leq \mu_{x,y} \leq 180^{\circ}$, then the
lens strengths remain bounded and satisfy the inequality

\noindent
\begin{eqnarray}
1\,\leq\,
\frac{|g_m|}{g}
\,\leq\, \frac{7 \sqrt{3} + 2}{6 \sqrt{3}}
\,\approx\,1.359,
\label{FODO_31}
\end{eqnarray}

\noindent
which, in particular, means that in this phase advance region
none of the lenses change
its polarity in comparison with their original FODO settings.
If, in addition, the scan of only $\mu_x$ is required with $\mu_y = 0$, then
the inequality (\ref{FODO_31}) is further relaxed to the
inequality

\noindent
\begin{eqnarray}
1\,\leq\,
\frac{|g_m|}{g}
\,\leq\, \frac{28 \sqrt{3} + 7}{24 \sqrt{3}}
\,\approx\,1.335,
\label{FODO_42}
\end{eqnarray}

\noindent
for $m = 1, 5, 9$, and to the inequality

\noindent
\begin{eqnarray}
1\,\leq\,
\frac{|g_m|}{g}
\,\leq\, \frac{24 \sqrt{3} + 1}{24 \sqrt{3}}
\,\approx\,1.024,
\label{FODO_43}
\end{eqnarray}

\noindent
for $m = 4, 8, 12$.

\begin{figure}[!htb]
    \centering
    \includegraphics*[width=85mm]{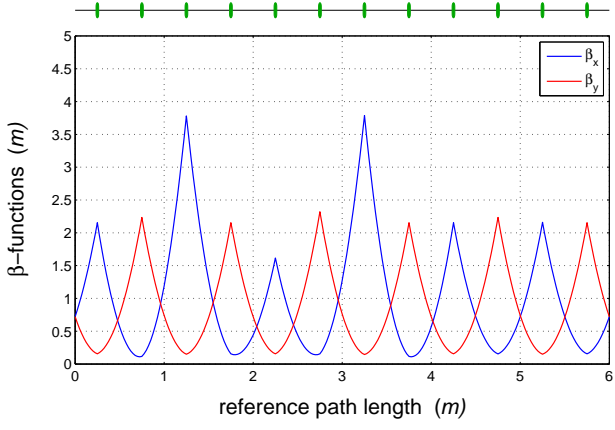}
    \caption{Betatron functions 
    along the phase advance scan beam line
    for $\mu_x = 60^{\circ}$, $\mu_y = 0^{\circ}$.}
    \label{fig1}
\end{figure}

\begin{figure}[!htb]
    \centering
    \includegraphics*[width=85mm]{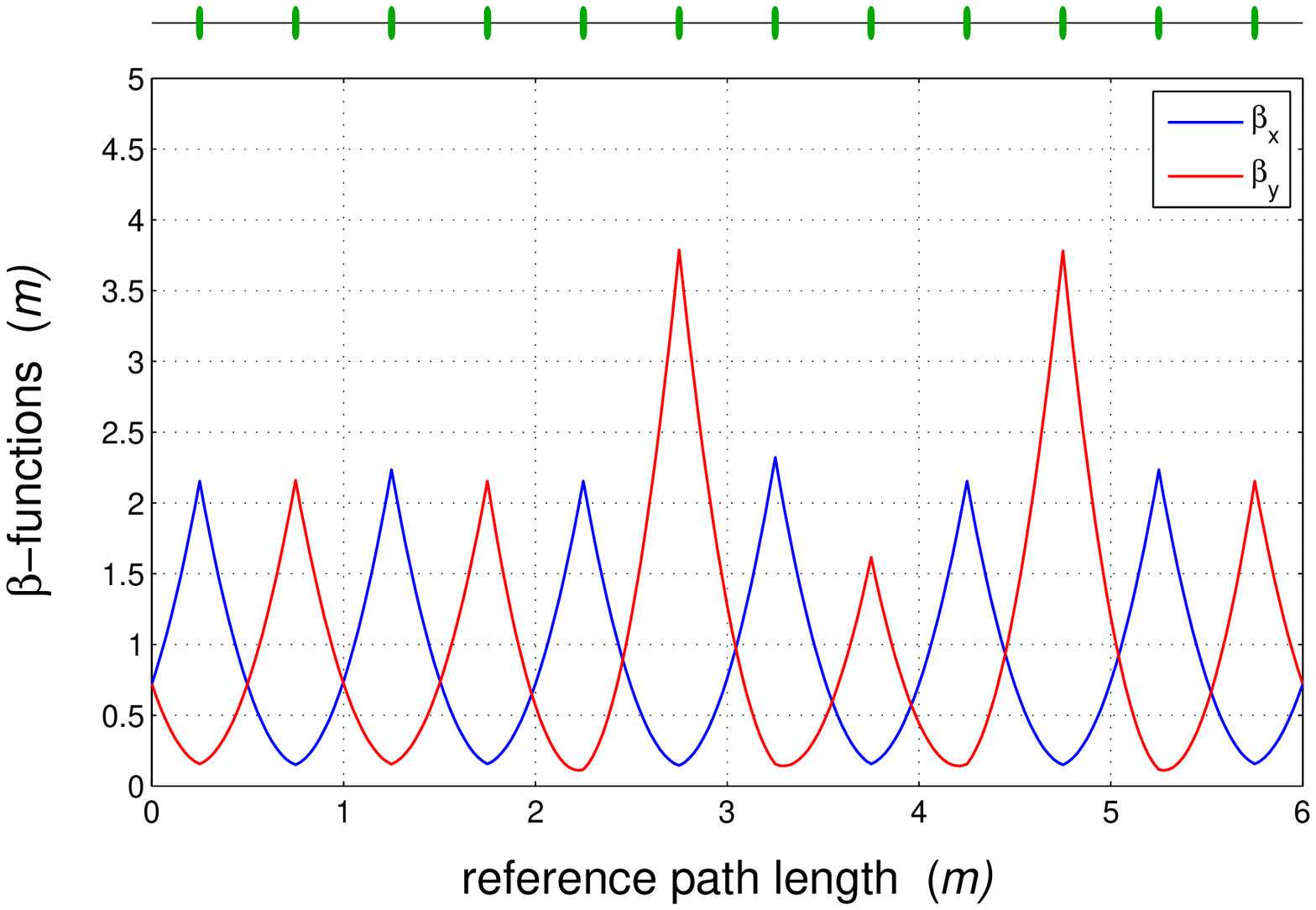}
    \caption{Betatron functions 
    along the phase advance scan beam line
    for $\mu_x = 0^{\circ}$, $\mu_y = 60^{\circ}$.}
    \label{fig2}
\end{figure}

\begin{figure}[!htb]
    \centering
    \includegraphics*[width=85mm]{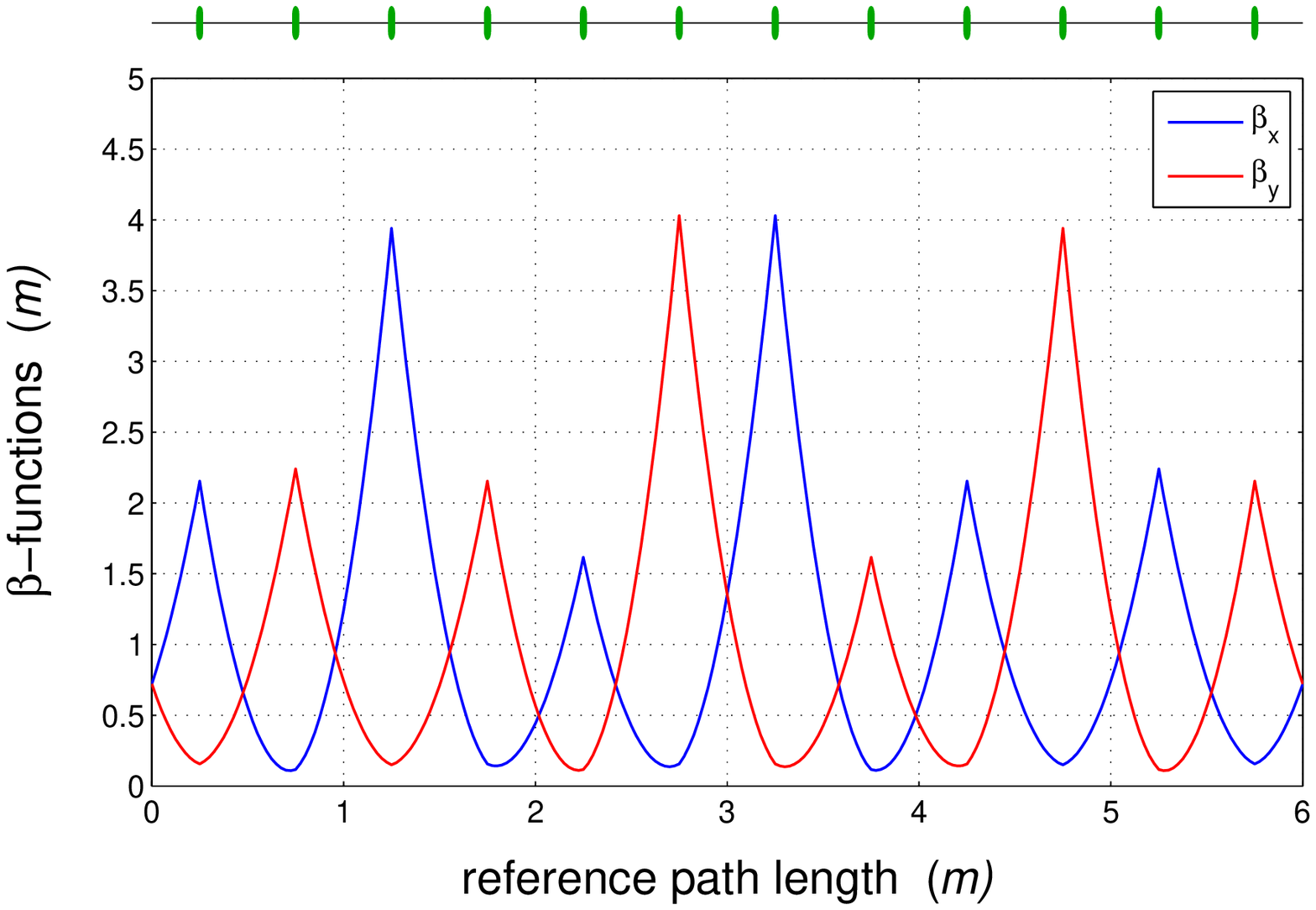}
    \caption{Betatron functions 
    along the phase advance scan beam line
    for $\mu_x = \mu_y = 60^{\circ}$.}
    \label{fig3}
\end{figure}

\begin{figure}[!htb]
    \centering
    \includegraphics*[width=85mm]{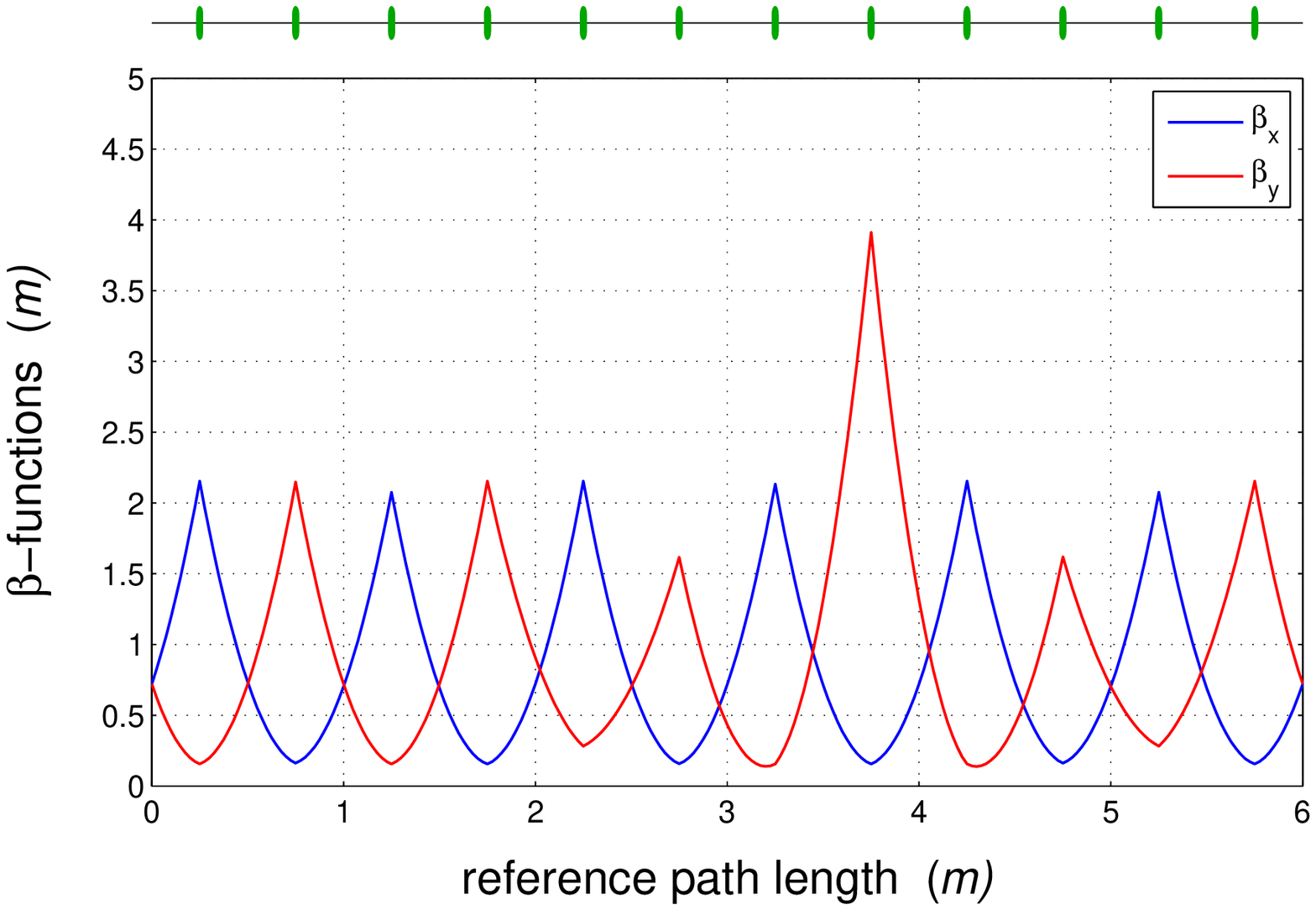}
    \caption{Betatron functions 
    along the phase advance scan beam line
    for $\mu_x = 0^{\circ}$, $\mu_y = -40^{\circ}$.}
    \label{fig4}
\end{figure}

\begin{figure}[!htb]
    \centering
    \includegraphics*[width=85mm]{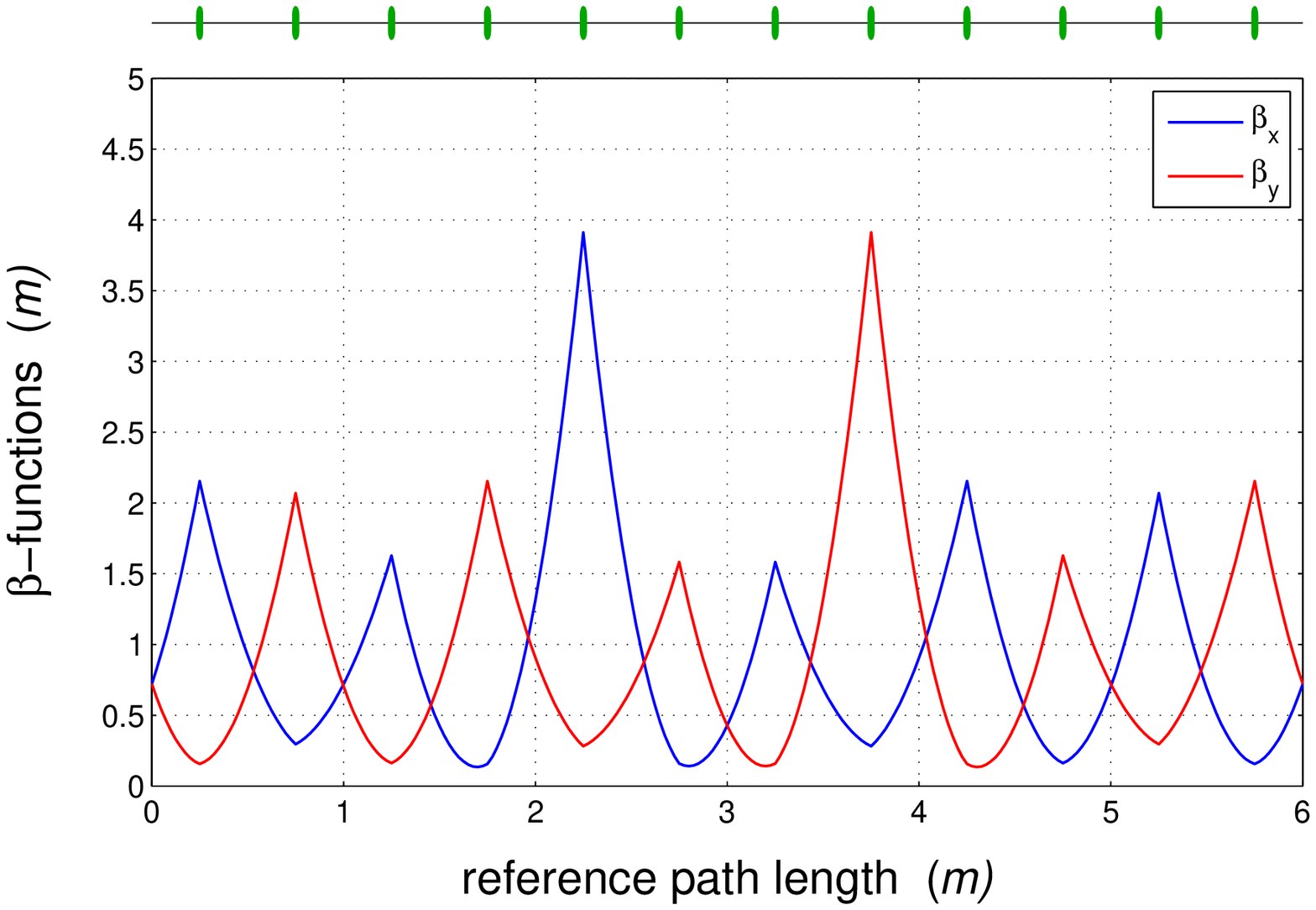}
    \caption{Betatron functions 
    along the phase advance scan beam line
    for $\mu_x = \mu_y = -40^{\circ}$.}
    \label{fig5}
\end{figure}

\begin{figure}[!htb]
    \centering
    \includegraphics*[width=85mm]{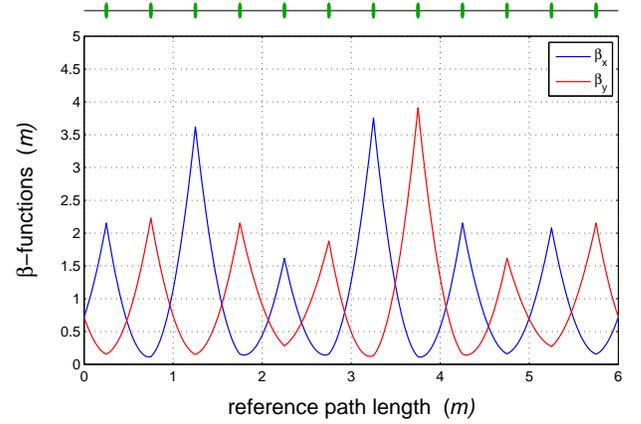}
    \caption{Betatron functions 
    along the phase advance scan beam line
    for $\mu_x = 60^{\circ}$, $\mu_y = -40^{\circ}$.}
    \label{fig6}
\end{figure}

Concerning the changes in the behavior of the betatron functions 
inside the beam line during the phase scan, then, for example, for all 
$\,-40^{\circ} \leq \mu_{x,y} \leq 60^{\circ}\,$ the betatron functions
$\beta_{x,y}(s)$ in each position $s$ along the beam line satisfy the inequalities

\noindent
\begin{eqnarray}
0.75 \, \beta_{min} \,\leq\,
\beta_{x,y}(s)
\,\leq\, 1.88 \, \beta_{max},
\label{FODO_47}
\end{eqnarray}

\noindent
where 

\noindent
\begin{eqnarray}
\beta_{min} \,=\,
\frac{2 - \sqrt{3}}{\sqrt{3}} \,L,
\;\;\;\;
\beta_{max} \,=\,
\frac{2 + \sqrt{3}}{\sqrt{3}} \,L
\label{FODO_48}
\end{eqnarray}

\noindent
are the minimum and the maximum of the periodic FODO solution.
In more details the behavior of the betatron functions
along the beam line can be seen in Figs. \ref{fig1}-\ref{fig6},
where they are drawn for the several values of
$\mu_{x,y}$ taken on the borders of the considered area and
for the FODO cell length chosen to be one meter.

The presented beam line for the scan of the phase
advances is simple, rather elegant and, in the same time,
can cover quite a range of phase advances
with not very large changes in the lens strengths as
compared to their original FODO settings.
It can also be adopted to the needs of the
European XFEL, where the linac between the two bunch compressors 
has exactly six FODO cells and two additional quadrupole groups
(matching sections) are available at both linac ends.

Note that it is not necessary to keep the periodic
matching conditions (\ref{FODO_1_1}). Any two sets of Twiss
parameters can be fixed at the beam line ends, but one has to
remember that the choice of them will affect the position of 
singularities of the solution obtained with the help of the 
three blocks with decoupled transverse actions. To avoid 
singularities completely and/or to have additional knob for
the control of the betatron functions inside the beam line,
one can switch to the solution
which utilizes four blocks or to the solution with
three blocks plus one additional lens. 
As described in Sec. IV of this paper,
the equal spacing of lenses can also be abandoned,
if required.

\end{document}